
\hoffset=0cm
\voffset=0cm
\hsize=16.0 true cm
\vsize=22.0 true cm
\parskip=\medskipamount
%
\font\mfiverm=cmr12 scaled\magstep5
\font\mfourrm=cmr12 scaled\magstep4
\font\mthreerm=cmr12 scaled\magstep3
\font\mfivesy=cmsy10 scaled 3583
\font\mfoursy=cmsy10 scaled 2986
\font\mthreesy=cmsy10 scaled 2488
\font\mfivebf=cmbx12 scaled 3583
\font\mfourbf=cmbx12 scaled 2986
\font\mthreebf=cmbx12 scaled 2488
\font\mfiveex=cmex10 scaled 3583
\font\mfourex=cmex10 scaled 2986
\font\mthreeex=cmex10 scaled 2488
\font\mfivei=cmmi12 scaled 3583
\font\mfouri=cmmi12 scaled 2986
\font\mthreei=cmmi12 scaled 2488





 \font\twelverm=cmr12     \font\twelvei=cmmi12
 \font\twelvesy=cmsy10 scaled 1200  \font\twelveex=cmex10 scaled 1200
 \font\twelvebf=cmbx12   \font\twelvesl=cmsl12
 \font\twelvett=cmtt12  \font\twelveit=cmti12

 \skewchar\twelvei='177   \skewchar\twelvesy='60


 \def\bigfont{\spaceskip=10 pt
   \normalbaselineskip=35pt
   \abovedisplayskip 12.4pt plus 3pt minus 9pt
   \belowdisplayskip 12.4pt plus 3pt minus 9pt
   \abovedisplayshortskip 0pt plus 3pt
   \belowdisplayshortskip 7.2pt plus 3pt minus 4pt
   \smallskipamount=3.6pt plus1.2pt minus1.2pt
   \medskipamount=7.2pt plus2.4pt minus2.4pt
   \bigskipamount=14.4pt plus4.8pt minus4.8pt
   \def\rm{\fam0\mfiverm}          \def\it{\fam\itfam\mfivei}%
   \def\sl{\fam\slfam\mfivesl}     \def\bf{\fam\bffam\mfivebf}%
   \def\mit{\fam 1}                 \def\cal{\fam 2}%
   \def\tt{\fourteentt}
   \textfont0=\mfiverm   \scriptfont0=\mfourrm   \scriptscriptfont0=\mthreerm
   \textfont1=\mfivei    \scriptfont1=\mfouri    \scriptscriptfont1=\mthreei
   \textfont2=\mfivesy   \scriptfont2=\mfoursy   \scriptscriptfont2=\mthreesy
   \textfont3=\mfiveex   \scriptfont3=\mfourex  \scriptscriptfont3=\mthreeex
   \textfont\itfam=\mfivei
   \textfont\slfam=\twelvesl
   \textfont\bffam=\mfivebf \scriptfont\bffam=\mfourbf
   \scriptscriptfont\bffam=\mthreebf
   \normalbaselines\rm}

 \def\twelvepoint{\normalbaselineskip=12.4pt
   \abovedisplayskip 12.4pt plus 3pt minus 9pt
   \belowdisplayskip 12.4pt plus 3pt minus 9pt
   \abovedisplayshortskip 0pt plus 3pt
   \belowdisplayshortskip 7.2pt plus 3pt minus 4pt
   \smallskipamount=3.6pt plus1.2pt minus1.2pt
   \medskipamount=7.2pt plus2.4pt minus2.4pt
   \bigskipamount=14.4pt plus4.8pt minus4.8pt
   \def\rm{\fam0\twelverm}          \def\it{\fam\itfam\twelveit}%
   \def\sl{\fam\slfam\twelvesl}     \def\bf{\fam\bffam\twelvebf}%
   \def\mit{\fam 1}                 \def\cal{\fam 2}%
   \def\tt{\twelvett}
   \textfont0=\twelverm   \scriptfont0=\tenrm   \scriptscriptfont0=\sevenrm
   \textfont1=\twelvei    \scriptfont1=\teni    \scriptscriptfont1=\seveni
   \textfont2=\twelvesy   \scriptfont2=\tensy   \scriptscriptfont2=\sevensy
   \textfont3=\twelveex   \scriptfont3=\twelveex  \scriptscriptfont3=\twelveex
   \textfont\itfam=\twelveit
   \textfont\slfam=\twelvesl
   \textfont\bffam=\twelvebf \scriptfont\bffam=\tenbf
   \scriptscriptfont\bffam=\sevenbf
   \normalbaselines\rm}



 \def\beginlinemode{\endmode
   \begingroup\parskip=0pt \obeylines\def\\{\par}\def\endmode{\par\endgroup}}
 \def\beginparmode{\endmode
   \begingroup \def\endmode{\par\endgroup}}
 \let\endmode=\par
 {\obeylines\gdef\
 {}}
 \def\singlespace{\baselineskip=\normalbaselineskip}
 
 \def\oneandahalfspace{\baselineskip=\normalbaselineskip
   \multiply\baselineskip by 3 \divide\baselineskip by 2}
 \def\doublespace{\baselineskip=\normalbaselineskip
\multiply\baselineskip by 2}

 \let\rawfootnote=\footnote              
 \def\footnote#1#2{{\rm\oneandahalfspace\parindent=0pt\parskip=12pt
   \rawfootnote{#1}{#2\hfill\vrule height 0pt depth 6pt width 0pt}}}
 \def\raggedcenter{\leftskip=4em plus 12em \rightskip=\leftskip
   \parindent=0pt \parfillskip=0pt \spaceskip=.3333em \xspaceskip=.5em
   \pretolerance=9999 \tolerance=9999
   \hyphenpenalty=9999 \exhyphenpenalty=9999 }

 \def\\{\cr}
 \twelvepoint            
 \oneandahalfspace         

 \overfullrule=0pt  

 \def\title                      
   {\null\vskip 3pt plus 0.2fill
    \beginlinemode \doublespace \myraggedcenter \bigfont}

 \def\author                     
   {\vskip 3pt plus 0.2fill \beginlinemode
    \singlespace \raggedcenter}

 \def\affil                      
   {\vskip 3pt plus 0.1fill \beginlinemode
    \oneandahalfspace
   \raggedcenter \sl}

 \def\abstract                   
   {\vskip 3pt plus 0.3fill \beginparmode
    \oneandahalfspace
    \centerline{\bf Abstract}

                                 }

 \def\endtopmatter               
   {\endpage                     
    \body}

 \def\body                       
   {\beginparmode}               

 \def\beneathrel#1\under#2{\mathrel{\mathop{#2}\limits_{#1}}}

 \def\refto#1{$[#1]$}           

 \gdef\refis#1{\item{#1.\ }}                     

 \def\figurecaptions             
   {\endpage
    \beginparmode
    \head{Figure Captions}
 }

 \def\endpage                    
   {\vfill\eject}

 \def\endpaper                   
   {\endmode\vfill\supereject}
 
 \def\endit
   {\endpaper\end}


 \def\heading                            
   {\vskip 0.5truein plus 0.1truein      
    \beginparmode \def\\{\par} \parskip=0pt \singlespace \raggedcenter}

 \def\subheading                         
   {\vskip 0.25truein plus 0.1truein     
    \beginlinemode \singlespace \parskip=0pt \def\\{\par}\raggedcenter}

 \def\tag#1$${\eqno(#1)$$}

 \def\align#1$${\eqalign{#1}$$}

 \def\aligntag#1$${\gdef\tag##1\\{&(##1)\cr}\eqalignno{#1\\}$$
   \gdef\tag##1$${\eqno(##1)$$}}

 \def\endaligntag{}

 \def\overset#1\to#2{{\mathop{#2}^{#1}}}
 \def\underset#1\to#2{{\mathop{#2}_{#1}}}

 \def\endreferences{\body}


 \def\ref#1{Ref.~#1}                     
 \def\Ref#1{Ref.~#1}                     
 \def\[#1]{[\cite{#1}]}
 \def\cite#1{{#1}}
 \def\(#1){(\call{#1})}
 \def\call#1{{#1}}
 \def\taghead#1{}
 \def\12{{1\over2}}

 \def\ie{{\it i.e.,\ }}
 
 \def\etal{{\it et al.\ }}

 \def\leaderfill{\leaders\hbox to 1em{\hss.\hss}\hfill}
 \def\twiddle{\lower.9ex\rlap{$\kern-.1em\scriptstyle\sim$}}
 \def\bigtwiddle{\lower1.ex\rlap{$\sim$}}
 \def\gtwid{\mathrel{\raise.3ex\hbox{$>$\kern-.75em\lower1ex\hbox{$\sim$}}}}
 \def\ltwid{\mathrel{\raise.3ex\hbox{$<$\kern-.75em\lower1ex\hbox{$\sim$}}}}
 \def\square{\kern1pt
\vbox{\hrule height 1.2pt\hbox{\vrule width 1.2pt\hskip 3pt
    \vbox{\vskip 6pt}\hskip 3pt\vrule width 0.6pt}
\hrule height 0.6pt}\kern1pt}
 \def\tdot#1{\mathord{\mathop{#1}\limits^{\kern2pt\ldots}}}
 
 \def\pmb#1{\setbox0=\hbox{#1}%
   \kern-.025em\copy0\kern-\wd0
   \kern  .05em\copy0\kern-\wd0
   \kern-.025em\raise.0433em\box0 }




 \def\references          
   {\headnn{References}
    \beginparmode                  
    \frenchspacing \parindent=0pt \leftskip=1truecm
    \parskip=8pt plus 3pt \everypar{\hangindent=\parindent}}

\catcode`@=11
\newcount\r@fcount \r@fcount=0
\newcount\r@fcurr
\immediate\newwrite\reffile
\newif\ifr@ffile\r@ffilefalse
\def\w@rnwrite#1{\ifr@ffile\immediate\write\reffile{#1}\fi\message{#1}}

\def\writer@f#1>>{}
\def\referencefile{
\r@ffiletrue\immediate\openout\reffile=\jobname.ref
  \def\writer@f##1>>{\ifr@ffile\immediate\write\reffile%
    {\noexpand\refis{##1} = \csname r@fnum##1\endcsname = %
     \expandafter\expandafter\expandafter\strip@t\expandafter%
     \meaning\csname r@ftext\csname r@fnum##1\endcsname\endcsname}\fi}%
  \def\strip@t##1>>{}}

\def\citeall#1{\xdef#1##1{#1{\noexpand\cite{##1}}}}
\def\cite#1{\each@rg\citer@nge{#1}} 

\def\each@rg#1#2{{\let\thecsname=#1\expandafter\first@rg#2,\end,}}
\def\first@rg#1,{\thecsname{#1}\apply@rg} 
\def\apply@rg#1,{\ifx\end#1\let\next=\relax
\else,\thecsname{#1}\let\next=\apply@rg\fi\next}

\def\citer@nge#1{\citedor@nge#1-\end-} 
\def\citer@ngeat#1\end-{#1}
\def\citedor@nge#1-#2-{\ifx\end#2\r@featspace#1 
  \else\citel@@p{#1}{#2}\citer@ngeat\fi} 
\def\citel@@p#1#2{\ifnum#1>#2{
\errmessage{Reference range #1-#2\space is bad.}%
\errhelp{If you cite a series of references by the notation M-N, then M and
    N must be integers, and N must be greater than or equal to M.}}\else%
 {\count0=#1\count1=#2\advance\count1 by1
\relax\expandafter\r@fcite\the\count0,%
  \loop\advance\count0 by1\relax
    \ifnum\count0<\count1,\expandafter\r@fcite\the\count0,%
  \repeat}\fi}

\def\r@featspace#1#2 {\r@fcite#1#2,} 
\def\r@fcite#1,{\ifuncit@d{#1}
    \newr@f{#1}%
    \expandafter\gdef\csname r@ftext\number\r@fcount\endcsname%
                     {\message{Reference #1 to be supplied.}%
                      \writer@f#1>>#1 to be supplied.\par}%
 \fi%
 \csname r@fnum#1\endcsname}
\def\ifuncit@d#1{\expandafter\ifx\csname r@fnum#1\endcsname\relax}%
\def\newr@f#1{\global\advance\r@fcount by1%
    \expandafter\xdef\csname r@fnum#1\endcsname{\number\r@fcount}}

\let\r@fis=\refis   
\def\refis#1#2#3\par{\ifuncit@d{#1}
   \newr@f{#1}%
   \w@rnwrite{Reference #1=\number\r@fcount\space is not cited up to now.}\fi%
  \expandafter\gdef\csname r@ftext\csname r@fnum#1\endcsname\endcsname%
  {\writer@f#1>>#2#3\par}}

\def\ignoreuncited{
   \def\refis##1##2##3\par{\ifuncit@d{##1}%
\else\expandafter\gdef\csname r@ftext\csname r@fnum##1\endcsname\endcsname%
     {\writer@f##1>>##2##3\par}\fi}}

\def\r@ferr{\endreferences\errmessage{I was expecting to see
\noexpand\endreferences before now;  I have inserted it here.}}
\let\r@ferences=\references
\def\references{\r@ferences\def\endmode{\r@ferr\par\endgroup}}

\let\endr@ferences=\endreferences
\def\endreferences{\r@fcurr=0
  {\loop\ifnum\r@fcurr<\r@fcount
\advance\r@fcurr by 1\relax\expandafter\r@fis\expandafter{\number\r@fcurr}%
    \csname r@ftext\number\r@fcurr\endcsname%
  \repeat}\gdef\r@ferr{}\endr@ferences}


\let\r@fend=\endpaper\gdef\endpaper{\ifr@ffile
\immediate\write16{Cross References written on []\jobname.REF.}\fi\r@fend}

\catcode`@=12

\citeall\refto  
\citeall\ref  %
\citeall\Ref  %

\ignoreuncited

\catcode`@=11
\newcount\tagnumber\tagnumber=0

\immediate\newwrite\eqnfile
\newif\if@qnfile\@qnfilefalse
\def\write@qn#1{}
\def\writenew@qn#1{}
\def\w@rnwrite#1{\write@qn{#1}\message{#1}}
\def\@rrwrite#1{\write@qn{#1}\errmessage{#1}}

\def\taghead#1{\gdef\t@ghead{#1}\global\tagnumber=0}
\def\t@ghead{}

\expandafter\def\csname @qnnum-3\endcsname
  {{\t@ghead\advance\tagnumber by -3\relax\number\tagnumber}}
\expandafter\def\csname @qnnum-2\endcsname
  {{\t@ghead\advance\tagnumber by -2\relax\number\tagnumber}}
\expandafter\def\csname @qnnum-1\endcsname
  {{\t@ghead\advance\tagnumber by -1\relax\number\tagnumber}}
\expandafter\def\csname @qnnum0\endcsname
  {\t@ghead\number\tagnumber}
\expandafter\def\csname @qnnum+1\endcsname
  {{\t@ghead\advance\tagnumber by 1\relax\number\tagnumber}}
\expandafter\def\csname @qnnum+2\endcsname
  {{\t@ghead\advance\tagnumber by 2\relax\number\tagnumber}}
\expandafter\def\csname @qnnum+3\endcsname
  {{\t@ghead\advance\tagnumber by 3\relax\number\tagnumber}}

\def\equationfile{%
  \@qnfiletrue\immediate\openout\eqnfile=\jobname.eqn%
  \def\write@qn##1{\if@qnfile\immediate\write\eqnfile{##1}\fi}
  \def\writenew@qn##1{\if@qnfile\immediate\write\eqnfile
    {\noexpand\tag{##1} = (\t@ghead\number\tagnumber)}\fi}
}

\def\callall#1{\xdef#1##1{#1{\noexpand\call{##1}}}}
\def\call#1{\each@rg\callr@nge{#1}}

\def\each@rg#1#2{{\let\thecsname=#1\expandafter\first@rg#2,\end,}}
\def\first@rg#1,{\thecsname{#1}\apply@rg}
\def\apply@rg#1,{\ifx\end#1\let\next=\relax%
\else,\thecsname{#1}\let\next=\apply@rg\fi\next}

\def\callr@nge#1{\calldor@nge#1-\end-}
\def\callr@ngeat#1\end-{#1}
\def\calldor@nge#1-#2-{\ifx\end#2\@qneatspace#1 %
  \else\calll@@p{#1}{#2}\callr@ngeat\fi}
\def\calll@@p#1#2{\ifnum#1>#2{\@rrwrite{Equation range #1-#2\space is bad.}
\errhelp{If you call a series of equations by the notation M-N, then M and
N must be integers, and N must be greater than or equal to M.}}\else%
 {\count0=#1\count1=#2\advance\count1 by1
\relax\expandafter\@qncall\the\count0,%
  \loop\advance\count0 by1\relax%
    \ifnum\count0<\count1,\expandafter\@qncall\the\count0,%
  \repeat}\fi}

\def\@qneatspace#1#2 {\@qncall#1#2,}
\def\@qncall#1,{\ifunc@lled{#1}{\def\next{#1}\ifx\next\empty\else
  \w@rnwrite{Equation number \noexpand\(>>#1<<) has not been defined yet.}
  >>#1<<\fi}\else\csname @qnnum#1\endcsname\fi}

\let\eqnono=\eqno
\def\eqno(#1){\tag#1}
\def\tag#1$${\eqnono(\displayt@g#1 )$$}

\def\aligntag#1\endaligntag
  $${\gdef\tag##1\\{&(##1 )\cr}\eqalignno{#1\\}$$
  \gdef\tag##1$${\eqnono(\displayt@g##1 )$$}}

\def\eqalignno#1{\displ@y \tabskip\centering
  \halign to\displaywidth{\hfil$\displaystyle{##}$\tabskip\z@skip
    &$\displaystyle{{}##}$\hfil\tabskip\centering
    &\llap{$\displayt@gpar##$}\tabskip\z@skip\crcr
    #1\crcr}}

\def\displayt@gpar(#1){(\displayt@g#1 )}

\def\displayt@g#1 {\rm\ifunc@lled{#1}\global\advance\tagnumber by1
        {\def\next{#1}\ifx\next\empty\else\expandafter
        \xdef\csname @qnnum#1\endcsname{\t@ghead\number\tagnumber}\fi}%
  \writenew@qn{#1}\t@ghead\number\tagnumber\else
        {\edef\next{\t@ghead\number\tagnumber}%
        \expandafter\ifx\csname @qnnum#1\endcsname\next\else
        \w@rnwrite{Equation \noexpand\tag{#1} is a duplicate number.}\fi}%
  \csname @qnnum#1\endcsname\fi}

\def\ifunc@lled#1{\expandafter\ifx\csname @qnnum#1\endcsname\relax}

\let\@qnend=\end\gdef\end{\if@qnfile
\immediate\write16{Equation numbers written on []\jobname.EQN.}\fi\@qnend}

\catcode`@=12

\catcode`@=11 
\def\lsim{\mathrel{\mathpalette\@versim<}}
\def\gsim{\mathrel{\mathpalette\@versim>}}
\def\@versim#1#2{\lower0.2ex\vbox{\baselineskip\z@skip\lineskip\z@skip
  \lineskiplimit\z@\ialign{$\m@th#1\hfil##\hfil$\crcr#2\crcr\sim\crcr}}}
\catcode`@=12 

\def\E#1 {\times 10^{#1}}
\def\pagenumbers{\pageno=1\global\footline={\hfil\twelverm-- \folio\ --\hfil}}

\newbox\charbox
\newbox\slabox
\def\sla#1{{\setbox\charbox=\hbox{$#1$}
           \setbox\slabox=\hbox{$/$}
           \dimen\charbox=\ht\slabox
           \advance\dimen\charbox by -\dp\slabox
           \advance\dimen\charbox by -\ht\charbox
           \advance\dimen\charbox by \dp\charbox
           \divide\dimen\charbox by 2
           \raise-\dimen\charbox\hbox to \wd\charbox{\hss/\hss}
           \llap{$#1$}
          }}

\catcode`@=11 

\newcount\chapterno
\global\chapterno=0
\newcount\appendixno
\global\appendixno=0
\newcount\headno
\global\headno=0
\newcount\subheadno
\global\subheadno=0
\newcount\figno
\global\figno=0
\newcount\tabno
\global\tabno=0
\newcount\noheadline
\newcount\curr@head
\newcount\c@ntentsno
\global\c@ntentsno=0
\newcount\str@pno
\newcount\c@ntents
\global\c@ntents=0
\newcount\currc@nt

\def\nolet#1{\ifcase#1{This shouldn't happen}\or A\or B\or C\or D%
\or E\or F\or G\or H\or I\or J\or K\or L\or M\or N\or O\or P\or Q%
\or R\or S\or T\or U\or V\or W\or X\or Y\or Z\else{\Omega}\fi}

\gdef\cano{}

\headline={\hfil}

\footline={\hfil\twelverm-- \folio\ --\hfil}

\gdef\s@s{ }

\output={
 \shipout\vbox{\columnbox}
 \advancepageno
\ifnum\outputpenalty>-20000 \else\dosupereject\fi}

\def\columnbox{\vbox{
\ifnum\firstmark=0\else
\ifnum\topmark=\botmark
\else\curr@head=\firstmark
\loop
\ifnum\curr@head>\c@ntents
\else
\ifnum\pageno=\csname c@ntentpage\number\curr@head\endcsname\else
\message{ReTeX for a correct contents page and the road to heaven}\fi\fi
\edef\contype{\iftrue\csname h@name\number\curr@head\endcsname\fi}
\ifnum\curr@head <\botmark \advance\curr@head by 1
\repeat
\fi\fi
\leftline{\vbox{\makeheadline\pagebody\makefootline}}}}

\def\contentspage#1#2#3#4{
\global\advance\c@ntents by1
\ifnum #1=\c@ntents
\expandafter\gdef\csname c@ntenttype#1\endcsname{#2}
\expandafter\gdef\csname c@ntenttext#1\endcsname{#3}
\expandafter\gdef\csname c@ntentpage#1\endcsname{#4}
\else
\message{Your contents page is garbled (no.#1) ... for a happy life correct
it}
\global\advance\c@ntents by-1
\fi
}

\def\bdotfil{\leaders\hbox to 1.5em{\hss.\hss}\hfil}

\def\contents{\ifnum\c@ntents=0
\message{Eh up .... where's the Contents page got to ?}
\else
\bighead{Contents}
{\currc@nt=1\loop
\message{\number\currc@nt}
\edef\c@ntype{\csname c@ntenttype\number\currc@nt\endcsname}
\if\c@ntype A
 \vskip 1cm\goodbreak\noindent
 {\bf\csname c@ntenttext\number\currc@nt\endcsname}
 \hfill\csname c@ntentpage\number\currc@nt\endcsname\fi
\if\c@ntype B
 \vskip 1cm\goodbreak\noindent
 {\bf\csname c@ntenttext\number\currc@nt\endcsname}
 \bdotfil\csname c@ntentpage\number\currc@nt\endcsname\fi
\if\c@ntype C
 \vskip 1cm\goodbreak\noindent
 {\bf\csname c@ntenttext\number\currc@nt\endcsname}
 \hfill\csname c@ntentpage\number\currc@nt\endcsname\fi
\if\c@ntype H
 \noindent\hbox{\hskip 1em}
 \csname c@ntenttext\number\currc@nt\endcsname
 \bdotfil\csname c@ntentpage\number\currc@nt\endcsname\fi
\if\c@ntype S
 \noindent\hbox{\hskip 2em}
 \csname c@ntenttext\number\currc@nt\endcsname
 \bdotfil\csname c@ntentpage\number\currc@nt\endcsname\fi
\break
\ifnum\currc@nt <\c@ntents \advance\currc@nt by 1
\repeat}\fi
}

\mark{0}

\def\head#1{                    
  \global\advance\headno by1
  \global\advance\c@ntentsno by1
  \global\subheadno=0
  \global\tabno=0
  \goodbreak\vskip 0.5truein    
\expandafter\xdef\csname h@name\number\c@ntentsno\endcsname
   {H\noexpand\else \number\headno\ #1}
   \immediate\write16{\number\c@ntentsno :#1}
    \taghead{\number\headno .}
\goodbreak
\vbox{
    \leftline{\bf\number\headno\ #1}
    \nobreak\vskip 0.25truein\nobreak
     }
    \mark{\number\c@ntentsno}
}

\def\headnn#1{
\vskip 0.5truein\goodbreak
\vbox{
    \leftline{\bf #1}
    \nobreak\vskip 0.25truein\nobreak
     }
}
\def\subhead#1{                    
  \global\advance\subheadno by1
  \global\advance\c@ntentsno by1
   \vskip 0.25truein\goodbreak    
\expandafter\xdef\csname h@name\number\c@ntentsno\endcsname
   {S\noexpand\else \cano .\number\headno.\number\subheadno\ #1}
   \immediate\write16{\number\c@ntentsno :#1}
\goodbreak
\vbox{
    \leftline{\bf\cano .\number\headno .\number\subheadno \ #1}
    \nobreak\vskip 0.1truein\nobreak
     }
    \mark{\number\c@ntentsno}
}

\def\fig#1#2{
	\global\advance\figno by 1
	\expandafter\xdef\csname f@g#1\endcsname{\number\figno}
	{\par\myraggedcenter{\it Fig.}\number\figno{} #2\smallskip}
}

\def\rfig#1{\hbox{{\it Fig.}\csname f@g#1\endcsname}}

\def\figref#1#2{\expandafter\xdef\csname f@g#1\endcsname{#2}}

\def\table#1#2{
	\global\advance\tabno by 1
	\expandafter\xdef\csname t@b#1\endcsname{\number\figno}
	{\myraggedcenter{\it Table} \t@ghead\number\tabno{} #2\smallskip}
}

\def\rtable#1{\hbox{{\it Table} \csname t@b#1\endcsname}}

\def\tablref#1#2{\expandafter\xdef\csname t@b#1\endcsname{#2}}

\def\first#1#2{{\str@pno=#1\expandafter\str@p#2 \end{} }}
\def\str@p#1 {\ifx\end#1\let\next=\relax%
\else\advance\str@pno by -1
\ifnum\str@pno=-1$...$\def\next##1\end{}%
\else #1 \let\next=\str@p\fi\fi \next}

\catcode`@=12 

 \def\myraggedcenter{\leftskip=0em plus 12em \rightskip=\leftskip
   \parindent=0pt \parfillskip=0pt \spaceskip=.3333em \xspaceskip=.5em
   \pretolerance=9999 \tolerance=9999
   \hyphenpenalty=9999 \exhyphenpenalty=9999 }

%
%
\def\frac#1#2{{\scriptstyle#1\over\scriptstyle#2}}

\def\Tr#1{\hbox{Tr}\noexpand\left[#1\noexpand\right]}

\def\ddot#1#2{\thinspace #1\!\cdot\! #2\thinspace}

\def\LEPI/{\hbox{\rm LEP I}}
\def\LEPII/{\hbox{\rm LEP I$\!$I}}
\def\GeV{{\rm\, GeV}}

\def\MeV{{\rm\, MeV}}

\font\circle=lcircle10
\def\trcorner{\hbox{\raise 2.9pt\hbox{\circle\char10}\kern -10pt}}

\def\gammav{\gamma{\!\raise -0.5em\hbox{$\scriptstyle 5$}}}



\input epsf

\nopagenumbers

\hfill\vbox{\rm\hbox{MAD/PH/833}
\hbox{February 1995}}
\title
Exponentiation of soft photons in a process involving hard photons
\bigskip
\author
D.J. Summers%
\footnote{${}^1$}{Email address: summers @ phenxr.physics.wisc.edu}
\affil
Department of Physics,
University of Wisconsin -- Madison,
1150 University Avenue,
Madison,
WI 53706,
U.S.A.
\bigskip

\abstract

We present a simple method of removing the singularities
associated with soft
photon emission to all orders in perturbation theory through
exponentiation, while keeping a consistent description of hard photon
emission. We apply this method to the process $e^+ e^- \to \mu^+ \mu^-
+ n\, \gamma$ where we include both $Z^0$ and $\gamma$ exchange and
retain the muon mass dependence. The photonic radiation is allowed to
be radiated off any charged leg, and so we include all initial and
final state radiation, as well as all interference effects.
The effect of exponentiation is to suppress soft photon emission over
the cross-section you would obtain from working at strictly leading
order. We also show how one would extend the method to treat the
collinear singularity; and remove the associated leading mass
logarithms.

\bigskip

\noindent {\bf PACS:} 12.15.Lk 13.40.Ks 11.80.Fv 13.38.Dy
\endpage
\body

\pagenumbers
\head{Introduction}

Photonic radiation off charged particles plays a very important part
in the physics of high energy particle colliders. At LEP initial state
photonic radiation is responsible for shifting the $Z^0$ peak and
altering its measured width, and hence an accurate description of this
radiation is necessary to extract the $Z^0$ boson mass and width in a
meaningful way. At
hadron colliders photonic radiation often forms a background (or
indeed a signal) to new processes; as, for example, in the future
search for an intermediate mass Higgs boson; or the testing of
anomalous gauge boson couplings. Clearly it is important to understand
this radiation.

If we calculate this photonic radiation at tree level in
perturbation theory as the radiation becomes either soft, or collinear
to a massless charged object, we encounter logarithms in
$E_\gamma / \sqrt{s}$ and $\theta$ respectively. When these logarithms
become large the probability that additional photons are also
radiated becomes large; and so the tree level description
breaks down. The situation can be improved by going to higher orders
in perturbation theory, however this does not cure the problem. If we
work at next to leading order then we include 1 additional photon and
so this is an improvement over leading order, but when additional
photonic radiation becomes important then this next to leading order
description breaks down, clearly this happens at each finite order in
perturbation theory. The solution to this is to go to fully
infinite order in perturbation theory; and this is possible in the
soft and collinear limits as in these limits the matrix element can be
well approximated by the soft approximation and Altarelli--Parisi
splitting functions. This means that all the logarithms associated
with soft emission can be resummed which leads to an exponential
series \refto{YRIII,JW,YFS3,BHLUMI4}.
The leading logs or next--to--leading logs associated with
collinear emission can be resummed and this leads to a ``parton
distribution'' for the charged particle \refto{YRIII,Collinear}.
These two methods each
respectively give an excellent description of soft and collinear
radiation. The resummation of the soft logarithms can be framed as a
reordering of the terms of perturbation theory in rigours way as was
first done by Yennie Frautschi and Suura (YFS) \refto{YFS}. This method
provides an excellent description of both soft and hard radiation
simultaneously. However there are few Monte Carlo programs that
incorporate YFS exponentiation and the best
(YFS3\refto{YFS3},BHLUMI4.0\refto{BHLUMI4})
only describe 1 hard photon exactly (with a $2^{\rm nd}$ included
through the
leading log splitting function). On the other hand tree level Monte
Carlos at a fixed order perturbation theory can describe arbitrarily
many hard photons \refto{Hard,mmff}.
In this paper we take our previous tree level Monte
Carlo \refto{mmff} that calculated $e^+e^- \to \mu^+ \mu^- + n \gamma$
and exponentiate the soft photons. We use a less rigorous, but more
simple, form of exponentiation than full YFS exponentiation. Our form
of exponentiation is equivalent to YFS exponentiation except that the
effects of the virtual loop Feynman diagrams are not explicitly
included, but only added using an ad hoc method. We show how one would
go about resumming the collinear radiation to remove the large mass
logarithms that occur, however as our primary interest in this paper
is the interface between hard and soft radiation we make no attempt to
include this resumed collinear radiation in our Monte Carlo program.
Now when we calculate radiation at tree level we are forced to include
cuts that keep us separate from both the soft and collinear region so
we do not encounter the singularities that are present there; in fact,
as we have suggested above these cuts should be large enough that we
do not approach the areas of phase space where the soft and collinear
logarithms become large. Now in our work as we resum the soft
logarithms and remove the soft singularity, and so we are no longer required
to keep the cut that keeps us separate from the soft corner of phase
space, however we are forced to retain the cut that keeps us separate from
the collinear singularity as we have not dealt with the singularities
that are there. As with tree level calculations this cut should be
viewed as an experimental cut, that should be chosen large enough to
keep us away from the large collinear logarithms, that is imposed on
all photons.

\head{Exponentiation of soft photons with other hard photons present}

In any process in which charged particles are accelerated the soft
photon approximation tells us that soft photons are predominately
radiated off external legs, and that the matrix element for the
Feynman diagram with a soft photon radiated off external
leg $i$ with charge $e$ is,
$$
{\cal M}\ e\ {\ddot{\epsilon}{p_i} \over \ddot{k}{p_i}} \tag
$$
where ${\cal M}$ is the matrix element for the process without a
photon, and $\epsilon$ and $k$ are respectively the photon polarisation
vector and momentum.

So if the matrix element for the process,
$$
\emptyset \to e^+ (p_1)\ e^- (p_2)\ \mu^+ (p_3)\ \mu^- (p_4) + n\gamma
\tag eemmnf
$$
is ${\cal M}_n$ the matrix element squared for the process,
$$
\emptyset \to e^+ (p_1)\ e^- (p_2)\ \mu^+ (p_3)\ \mu^- (p_4)
     + n \gamma + \gamma_s (k)\tag
$$
in the soft photon approximation is given by,
$$
|{\cal M}_{n,1}|^2 = - e^2 |{\cal M}_n|^2
 \left( {p_1 \over \ddot{p_1}{k}} - {p_2 \over \ddot{p_2}{k}} +
          {p_3 \over \ddot{p_3}{k}} - {p_4 \over \ddot{p_4}{k}} \right)^2
\tag eikonal
$$
after we sum over the different spin states of the photon, where we
definite all particles in the final state. This means that initial
state particles will have negative energy.

If the soft photon and the $n$ original photons are in mutually
exclusive areas of phase space then there is no symmetry factor
between the two. We can force this to happen by only considering the
original $n$ photons with energy, $E_h$, larger than some cut, and soft
photons with energy, $E_s$, less than that cut, \ie
$$
E_h > E_{\rm cut} > E_s \tag
$$
then differential cross-section is given by,
$$
\eqalignno{
d \sigma_{n+1} &= - e^2 {1 \over \hbox{Flux}} |{\cal M}_n|^2
 \left( {p_1 \over \ddot{p_1}{k}} - {p_2 \over \ddot{p_2}{k}} +
             {p_3 \over \ddot{p_3}{k}} - {p_4 \over \ddot{p_4}{k}} \right)^2
   d (\hbox{LIPS})_{n+3} &() \cr
	&\simeq {1 \over \hbox{Flux}} |{\cal M}_n|^2 d (\hbox{LIPS})_{n+2}
 \quad (\hbox{Eikonal Factor})\
{E^2 dE\, d\Omega \over 2 E (2 \pi)^3} &() \cr}
$$
where,
$$
\eqalignno{
(\hbox{Eikonal Factor}) &= -e^2
	\left( {p_1 \over \ddot{p_1}{k}} - {p_2 \over \ddot{p_2}{k}} +
        {p_3 \over \ddot{p_3}{k}} - {p_4 \over \ddot{p_4}{k}} \right)^2 &()\cr
	&\equiv e^2 { f(\Omega) \over E^2 } &() \cr}
$$
and integrating over the soft photon phase space for photon energies
$E_{\rm min} < E_\gamma < E_{\rm cut}$ gives,
$$
d \sigma_{n,1} = \underbrace{{1 \over \hbox{Flux}}\, |{\cal M}_n|^2
             \, d (\hbox{LIPS})_{n+2}}_{\textstyle = d \sigma_n} \quad
	      {\alpha \over 4 \pi^2}\,
 \ln \left( {E_{\rm cut} \over E_{\rm min}} \right) \, g(\Omega_c)
\tag sigma1
$$
with,
\footnote{${}^2$}{We give the explicit form of $g$ in the appendix.}
$$
g(\Omega_c)=\int_{\Omega_c} f(\Omega) d \Omega \tag
$$
and so we see from \(sigma1) that the probability to emit a soft
photon just factorises the lower order cross-section. However in the
factorising term we have a contribution from $\ln(E_{\rm min})$ and this
diverges as $E_{\rm min}$  goes to zero,
this is the soft singularity.\footnote{${}^3$}
{Also note that the term $g(\Omega_c)$ diverges in the
limit of massless fermions when the photon becomes collinear to any
fermion. This is the collinear singularity, isolating this singularity
and removing it leads to collinear description of photon radiation. In
this work we will apply angular cuts to {\it all} photons (be they
hard or soft) to keep the photons separate from the
fermions, and as such we work in an area of phase space where these
collinear singularities are not important.}

Now for fixed particle momenta in $d\sigma_n$ unitarity tells us that
when we integrate over the soft photon momenta that $d\sigma_{n,1}$
is finite. As the soft, non colinear,
singularity contains no electron or muon mass terms this tells us that
the soft logarithm must cancel in the full calculation including
virtual diagrams\refto{BN}.
We can achieve the same effect as the virtual diagrams by imposing
an effective lower energy cut off on the photon energy, $E_{\rm reg}$,
and for the higher order corrections corrections to be
${\cal O}(\alpha )$ we require,
$$
\ln \left( {\sqrt{s}/2 \over E_{\rm reg} }\right) = {\cal O}(1) \tag Ereg
$$
where $\sqrt{s} / 2$ is the maximum photon energy.
So the cross-section, differential in $n$ photons with energy larger
than $E_{\rm cut}$ and integrated over 1 soft photon with energy
smaller than $E_{\rm cut}$ is given by,
$$
d \sigma_{n,1} = d \sigma_n \quad
	      {\alpha \over 4 \pi^2}\,
 \ln \left( {E_{\rm cut} \over E_{\rm reg}} \right) \, g(\Omega_c)
\tag
$$
In doing this we improve the accuracy of the differential
cross-section calculation from
${\cal O}( \alpha \ln(E))$ to ${\cal O}( \alpha)$.
The lack of knowledge of $E_{\rm reg}$ beyond \(Ereg) is
exactly the lack of knowledge that we have from doing a tree level
calculation without calculating the virtual diagrams.
Also note that if we consider the differential cross-section
for values of $E_{\rm cut} < E_{\rm reg}$ then this
appears to go negative ! This is just an artifact of working, at least
as far as the soft logs go, beyond tree level.

If we now consider additional soft photonic radiation then for the process,
$$
\emptyset \to e^+ (p_1)\ e^- (p_2)\ \mu^+ (p_3)\ \mu^- (p_4)
+ n \gamma + \gamma_s (k_1) \cdots \gamma_s (k_m)\tag
$$
then the soft photon approximation gives the matrix element squared
as,
$$
|{\cal M}_{n,m}|^2 = |{\cal M}_n|^2
\prod_{i=1}^m  - e^2  \left( {p_1 \over \ddot{p_1}{k_i}}
- {p_2 \over \ddot{p_2}{k_i}} +
      {p_3 \over \ddot{p_3}{k_i}} - {p_4 \over \ddot{p_4}{k_i}} \right)^2
\tag
$$
The phase space for these soft photons also approximately factorises,
$$
d(\hbox{LIPS})_{2+n+m} \simeq d(\hbox{LIPS})_{2+n} \
\prod_{i=1}^m {E_i^2 dE_i\, d\Omega_i \over 2 E_i (2 \pi)^3}
\tag
$$
and so as before we can write the differential cross-section for the
process with $n$ photons with energy larger than $E_{\rm cut}$ and
integrated over $m$ soft photons with energy less that $E_{\rm cut}$ as,
$$
d\sigma_{n,m} = d \sigma_n {1 \over m!}
\left( {\alpha \over 4 \pi^2}
   \ln \left( {E_{\rm cut} \over E_{\rm reg}} \right)
g(\Omega_c) \right) ^m
\tag
$$
where the $1 / m!$ term is the symmetry factor that we get from
integrating the $m$ identical soft photons over the same regions of phase
space.

If we now ask what the differnetial cross-section,
$d\sigma_n^{\rm exp}$, for process where we
produce an arbitary number of photons satisfying the cut $\Omega_c$,
where we stay differential in $n$ photons with energy larger than
$E_{\rm cut}$ and integrate over an arbitary number of
photons with energy less than
$E_{\rm cut}$, we find,
$$
\eqalignno{
d\sigma_n^{\rm exp} &= d\sigma_{n,0} + d\sigma_{n,1} + d\sigma_{n,2}
                       + \ldots &()\cr
&= d\sigma_n \left( 1 +
    \left({\alpha \over 4 \pi^2}
      \ln \left( {E_{\rm cut} \over E_{\rm reg}} \right)g(\Omega_c) \right)
  + {1 \over 2!} \left({\alpha \over 4 \pi^2}
  \ln \left( {E_{\rm cut} \over E_{\rm reg}} \right)g(\Omega_c) \right)^2
  + \ldots \right) \qquad&() \cr
&= d\sigma_n \ \exp \left({\alpha \over 4 \pi^2}
      \ln \left( {E_{\rm cut} \over E_{\rm reg}} \right)
	g(\Omega_c) \right) &()\cr
&= d\sigma_n \ \left({E_{\rm cut} \over E_{\rm reg}}\right)^{\left(\alpha
      g(\Omega_c) / 4 \pi^2 \right)} &(exp0)\cr
}$$
$d\sigma^{\rm exp}_n$ has the sum over an infinite number of soft
photons, and this means that it has all the real soft logarithms
resummed in it, each with the soft singularity removed; this means it
will provide an accurate description
of arbitrarily soft photons. Notice that in this ``exponentiated''
form we have no problems with cross-sections going negative for values
of $E_{\rm cut} < E_{\rm reg}$.
$d\sigma^{\rm exp}_n$ has no knowledge of photons that fail the
angular cut $\Omega_c$ and with respect to those photons
$d\sigma^{\rm exp}_n$ is a strictly leading order quanity. This means
$d\sigma^{\rm exp}_n$ does not describe this collinear radiation, and
in particular when large collinear logs appear
in $g(\Omega_c)$ or $d\sigma_n$,
$d\sigma^{\rm exp}_n$ will not give a good description of the radiation.

Now we can use the form of \(exp0) to arrive at a workable Monte Carlo
stratergy;

\item{$\bullet$} Choose the number of photons $n$ to be produced.

\item{$\bullet$} Calculate a point in phase space for the process
\(eemmnf) where the $n$ photons have energy larger than $E_{\rm cut}$.

\item{$\bullet$} Calculate the hard matrix element for the process
\(eemmnf) and weight it by,
$$
\left({E_{\rm cut} \over E_{\rm reg}}\right)^{\left(\alpha
      g(\Omega_c) / 4 \pi^2 \right)} \tag expfac
$$

\item{$\bullet$} Sum over the events to perform the integration over
$n$ and the hard phase space for the process \(eemmnf) .

Notice that we can only calculate the expression \(expfac) after
calculating the point in the hard phase space as the $g(\Omega_c)$
term depends upon the orientations of the charged particles in the
process.

In this prescription we appear to have introduced two extra
parameters, $E_{\rm reg}$ and $E_{\rm cut}$, over the tree level
calculation. Now the value of $E_{\rm reg}$ is introduced to cancel
the soft singularity; and as in a tree level calculation we don't
know how much of
the singularity cancels, the value of $E_{\rm reg}$ is unknown beyond
what we learn from equation \(Ereg). Thus the dependence of our
calculation on $E_{\rm reg}$ is a measure of our uncertainty in only
calculating the real photon emission diagrams without calculating the
virtual photon loop diagrams that cancel the soft singularities. The
same level of uncertaintly is also present at tree level although there
is no parameter that displays it.

Our apparent dependence upon $E_{\rm cut}$ is more worrying.
$E_{\rm cut}$ was just introduced as a parameter to distinguish the
``hard'' photons from the ``soft'' photons, however there is no
physical meaning to the words ``hard'' and ``soft''; there is no
magical energy where photons suddenly become ``hard''.
Consider what happens as we decrease $E_{\rm cut}$; for the hard
matrix elements we approach the soft singularity and the cross-section
grows rapidly; now this increase in cross-section comes about from low
energy photons, as these photons have low energy they do not
change the physical signature of the process. However as the hard
matrix elements are growing the eikonal factor \(expfac) decreases
and if the value of $E_{\rm cut}$ is to have no physical significance
then these two effects should cancel each other. This will give us a
very strong test of our results, if we have a result independent of
$E_{\rm cut}$ we are practically required to a correct calculation of
both the hard matrix element and the soft eikonal factor.
Notice that as we include an arbitrary number of soft photons in the
soft exponentiating factor we are required to consider an arbitrary
number of hard photons in order for the $E_{\rm cut}$ cancellation to
occur. This is true even if those hard photons would not be
experimentally observed on energy grounds alone. This is
the opposite of what we do in a strictly LO calculation; there if we
require say the 2 photon differential cross-section it is essential that
we generate exactly 2 photons, here it is essential that
we also calculate the $n$ photon rate for all $n \geq 2$. Also notice
that we will only get this cancelation between the soft photon
exponentiation and the hard photon cross-section if we treat both the
soft and hard photons identically, in particular this means that the
angular cuts that we apply to the hard photons must also be applied to
the soft photons.

Comparing the naive form of exponentiation considered here with
the more rigorous YFS exponentiation, then YFS exponentiation
explictly includes both hard photon corrections and virtual photon
corrections to the exponentiating factor.
The virtual photon corrections are totally ignored in this
work, as we consider no virtual diagrams; this means that this work
has uncertainties of ${\cal O}(\alpha) \approx 1\%$ -- these
uncertainties manifest themselves as a lack of knowledge in the value
of $E_{\rm reg}$. The hard photon corrections give only a finite
contribution to the matrix element in the soft corner of phase space,
and as the volume of this phase space tends to zero as
$E_{\rm cut} \to 0$ in this limit the hard photon corrections have
vanishing effect. In practice this means that we should take
$E_{\rm cut}$ small enough that our results have no dependance upon
$E_{\rm cut}$. In YFS exponentiation the total energy of all radiated
soft photons is forced to be less than some cut; whereas in our form
of exponentiation it is each soft photon energy individually that has
energy less than some cut; this means that there is a worry that
although each soft photon individually only carries away a small
amount of energy, as there are in infinite number of soft photons they
may carry away a sizable amount of energy in total. If this were to
happen then our results would be incorrect because energy would not be
conserved in the $E_{\rm cut} \to 0$ limit; however if this were to
happen our results would not be independent of the cut $E_{\rm cut}$,
thus independance of our results upon this cut check that our energy
cut on the soft photons is valid.

\head{Collinear radiation}

In this paper we are primarily interested in the interaction between
soft and hard radiation, as such we have only included the effects of
resumming soft radiation in our Monte Carlo program. This means
that we are always forced to include an ``experimental''
angular cut on all photons, $\Omega_c$, that keeps us separate from
the singularity when radiation becomes collinear to a massless charged
object. As we have this ``experimental cut'' our method has nothing to
say about photons that fail the cut $\Omega_c$, or the physics that
these photons generate (like the shift in the measured $Z^0$ mass and
width).
In this section we show how such collinear radiation could
be included in our method at the leading collinear log level;
however we make no
attempt to include this collinear radiation in a Monte Carlo set up.

In principle we would like to proceed in a similar way to the soft
radiation, that is we evaluate the matrix element for a charged
particle radiating an arbitrary number of collinear photons, then
integrate the photon momenta over the collinear region of phase space;
and finally sum over an arbitrary number of collinear photons. In
practice the matrix element for a charged particle radiating an
arbitrary number of collinear photons is not known and so this method
fails. This means that we can not calculate all real collinear
logarithms in the way that we calculated all real soft logarithms;
however we can calculate the leading logs or the next to leading logs
through an evolution equation. If we have 2 collinear photons then if
one photon is far more collinear than the other we can approximate the
matrix element as two independent collinear emissions of photons.
Usually the evolution equation is in terms of the virtuality of the
initial of final charged particle, or the maximum $p_T$ that the
radiation can have; however for our case it is far more convenient for
the evolution to be done in terms of the angle of the emitted
radiation.

If we have an massless charged particle, $p$, that radiates a
collinear photon, $k$, at an angle $\theta$; so the final charged
particle, $p'$, has a fraction $z$ of the initial particle energy,
then the lowest order matrix element is multiplied by,
$$
|{\cal M}_{\rm split}|^2 = 2 e^2 {1\over p'.k} {\cal P}(z)
         = 2 e^2 {1\over E' E_\gamma (1-\cos\theta)} {\cal P}(z) \tag
$$
where,
$$
{\cal P}(z)= \left( {1+z^2 \over 1-z} \right)_+ \tag
$$
and in the collinear region the phase space is given by,
$$
d(\hbox{LIPS}) = {\pi \over 2}\ dz\ d(2p'.k)
	= \pi E' E_\gamma\ dz\ d\cos\theta \tag
$$
where we have integrated over the unimportant azimuthal angle.
So the differential cross-section to emit an extra single collinear
photon is given by,
$$
|{\cal M}_{split}|^2 d(\hbox{LIPS}) = (2 \pi e^2)\ d(-\ln(1-\cos\theta))
				\ {\cal P}(z) dz		\tag
$$
If we now consider the cross-section to emit an arbitrary number of
collinear photons up to some angle $\theta$ then the lowest order
cross-section is multiplied by a function, $D$; for this to be useful
we need to know the energy fraction of the final charged particle, $z$,
and so it is convenient to define,
$$
{\cal D}(z) = {d D \over dz} \qquad\hbox{where}\qquad
D = \int_0^1 dz\ {\cal D}(z)\tag
$$
then the evolution of ${\cal D}(z)$ is given by,
$$
{d {\cal D }(z) \over d(ln (1 - \cos\theta))}
	= (2\pi e^2) \int_z^1 {dx \over x} {\cal P}(x) {\cal D}(z/x)
\tag sDevol
$$
Where to derive this equation we have assumed that the splitting
function for several angular ordered photons is given by a production
of splitting functions, this is strictly only true when the photons
are strongly ordered. This means that this evolution equation only
sums the leading collinear logarithms.

If we integrate \(sDevol) over $z$ we find,
$$
D=\lambda\, (1-\cos\theta)^{2\pi e^2 P} \tag Dsol
$$
where $P=\int_0^1 dz\ {\cal P}(z)$. Usually within the + prescription
$P \equiv 0$, and so $D$ is constant when evolved in $\theta$.
This means that as we vary the angular cut arround the charged
particles $\Omega_c$ the cross-section in the collinear region does
not change. However the cross-section for the noncollinear radiation
does depend upon the angular cut $\Omega_c$, and this means that the
total cross-section is not independant of this cut. In order to restore
indepenence of the cross-section on $\Omega_c$ we define
$\tilde{\cal P}$ ,
$$
\tilde{\cal P}(z)= {\cal P}(z)
   + \delta(z-1) \int_0^{1-E_{\rm reg}/E} dy\ {\cal P}(y) \tag
$$
where $E$ is the energy of the charged particle, and then
use $\tilde{\cal P}$ in place of $\cal P$.

Equation \(sDevol) being an evolution equation
only tells us how radiation up to
some angle from the charged particle is related to radiation at any
other angle, it does not tell us where to start the evolution.
Now if
the charged particle were indeed massless then the starting point of
the evolution would be uncalulable, as in QCD parton distributions for
the proton and
fragmentation functions; however for massive charged particles
radiation is supressed in a dead cone surrounding the charged particle
of angle defined by $\sin\theta_{\rm DC} \simeq m/E$,
where $m$ and $E$ are
respectivly the mass and original energy of the charged particle. This
suggests the starting point of the evolution as,
$$
{\cal D}(z, \theta_{\rm DC}) = \delta(z-1) \tag
$$
This gives that,
$$
D(\theta_{\rm DC}) = 1 \tag
$$
\ie $\lambda\simeq \left({ 2 E^2 \over m^2}\right)^{2 \pi e^2 P} $,
this means that,
$$
D(\cos\theta \approx 0) \simeq
\left({ 2 E^2 \over m^2}\right)^{2 \pi e^2 P} \tag
$$
Now in the limit $m \to 0$, $D$ diverges; this, at least for final
state radiation, is unphysical\refto{BN}. The origin of this logarithmic
divergence is in the integral over $\cos\theta$ in the differential
cross-section, this takes the form,
$$
\int_{\cos\theta\approx 0}^{\cos\theta \approx \sqrt{ 1 - m^2/E^2 }}
 d(-\ln (1 - \cos\theta ) \simeq \ln \left({2E^2 \over m^2}\right)
\tag
$$
This is the collinear logarithm that, like the soft logarithm, also
cancels (for final state radiation \refto{KLN}) on the
virtual diagrams. As with
the soft singularity we have not calculated the virtual diagrams we
again do not know how much of the mass logarithm cancels, and so we
again introduce an extra parameter that quantifies our lack of
knowledge in this cancellation. For the mass logarithms to be absent
in the total cross-section we require,
$$
D(\cos\theta \approx 0) = 1 + {\cal O}(\alpha) \tag
$$
and for this to be true we choose $\lambda$ in equation \(Dsol) to have the
value,
$$
\lambda= 1 + \mu \alpha \qquad\hbox{ where }\qquad \mu\sim 1 \tag
$$
This means that,
$$
D(\theta_{\rm DC}) \simeq (1 + \mu \alpha)
		\left({m^2 \over 2E^2}\right)^{2\pi e^2 P} \tag
$$
and so we choose the starting point for the evolution \(sDevol) to be,
$$
{\cal D}(z,\theta_{\rm DC}) = \delta(z-1)\ (1 + \mu \alpha)\
		\left({m^2 \over 2E^2}\right)^{2\pi e^2 P} \tag
$$
This ensures that the total cross-section is not enhanced by
logarithms in $m$. For angles less that $\theta_{\rm DC}$ the
fragmentation function $\cal D$ no longer evolves in $\cos\theta$ but
stays at the value of ${\cal D}(\theta_{\rm DC})$.
It should be noted that we require a different
value for $\mu$ for initial and final state radiation; indeed whereas
for final state radiation we know that $\mu\sim 1$\refto{KLN},
we know no such thing for initial state radiation where the
cross-section may contain logarithms in the mass. Also note that
${\cal D}(z)$ plays a very different role for initial state radiation
than for final state radiation. For initial state radiation the
charged particle fragments into a charged particle and photons {\it
before} the hard scattering, and so the charged particle's momentum is
degraded and the hard scattering takes place at a lower $\sqrt s$;
whereas for final state radiation the fragmentation of the
charged particle happens after the hard scattering and so does not
change the hard scattering.

Having set up the fragmentation functions $D$ and $\cal D$ the hard
differential cross-section, where we have summed over an arbitary
number of soft and/or collinear photons, is given by,
$$
\eqalign{
d\sigma_n^{\rm col-exp} = & d\sigma_n \Big|_{z_1,z_2} \cr
&\quad\times \left({E_{\rm cut} \over E_{\rm reg}}\right)^{\left(\alpha
      g(\Omega_c) / 4 \pi^2 \right)} \cr
&\quad\times {\cal D}(\Omega_c,z_1)\,dz_1
\ {\cal D}(\Omega_c,z_2)\,dz_2
\ {\cal D}(\Omega_c,z_3)\,dz_3
\ {\cal D}(\Omega_c,z_4)\,dz_4 \cr}
\tag csdsigma
$$
where $d\sigma_n \Big|_{z_1,z_2}$ is the hard differential
cross-section where the incomming electrons have fractions $z_1$ and
$z_2$ of the beam energy; $z_3$ znd $z_4$ are the fractional
energy that the final state muons have.

As the soft radiation is radiated off
particles with a similar virtuality as the collinear radiation, or
alternativtly they are radiated at the same timescale, it is not clear
wether $g(\Omega_c)$ should be calculated before of after the
collinear radiation has been emitted in the $\cal D$ functions.
However recall that $g$ is the integral over the matrix element
squared, where the matrix element is the sum of terms that go like
$p / \ddot p k$; and this depends only on the direction of $p$ and not
its energy in the massless limit.
So $g$ only depends on the directions of the charged
particles, and not on the amount of energy they carry. This means that
$g$ is independent, in the massless collinear limit, of wether it is
calculated before or after collinear radiation is emitted.

With the differential cross-section in the form \(csdsigma) it is easy
to give a Monte Carlo mechanism for generating hard photons.

\item{$\bullet$} Choose the number of photons $n$ to be produced.

\item{$\bullet$} Choose $z_1$ and $z_2$ and hence the initial
electron energies.

\item{$\bullet$} Generate the hard phase space for $n$ photons where
the initial electrons have fraction $z_1$ and $z_2$ of the beam
energy.

\item{$\bullet$} Choose $z_3$ and $z_4$ that give the fractional
energy that the observed muons have over the muons that emerge from
the hard scattering.

\item{$\bullet$} Calculate the hard matrix element for hard phase
space and weight it by,
$$
\ \left({E_{\rm cut} \over E_{\rm reg}}\right)^{\left(\alpha
      g(\Omega_c) / 4 \pi^2 \right)}
\ {\cal D}(\Omega_c,z_1)
\ {\cal D}(\Omega_c,z_2)
\ {\cal D}(\Omega_c,z_3)
\ {\cal D}(\Omega_c,z_4)
$$

\item{$\bullet$} Sum over events to perform the integration over the
hard phase space, $z_1$ through $z_4$ and $n$ in the usual Monte Carlo
way.

In the soft case the dependance upon $E_{\rm cut}$ cancels exactly
between the hard cross-section and the soft exponential factor. In
this collinear case we do not expect exact cancelation of $\Omega_c$
dependance between the collinear factor $D$ and the hard
cross-section, as we have only resummed the leading logs in $D$ rather
than all the logs as in the soft case. Non the less there should only
be rather mild dependance on $\Omega_c$ arrising from the sub leading
logs that occur in the hard cross-section, but not in the factor $D$.

The resummation of these collinear logs has no analogy in YFS
exponentiation, which makes no attempt to resum collinear logs. Within
YFS exponentiation the full collinear logs can only be added as hard
corrections to the exponentiating factor. This means that our
collinear resummation should give a more accurate description of
radiation in the collinear region than YFS exponentiation.

\head{Numerical Results}

We have applied the procedure from section 2 to the process,
$$
e^+ e^- \to \mu^+ \mu^- + n\,\gamma \tag eemumupr
$$
for arbitrary $n$. To do this we need the hard matrix element for
\(eemumupr), this we have calculated in a previous paper \refto{mmff}.
In that calculation we retained the full mass dependence of the muons,
and allow an arbitrary number of hard photons radiated from both the
initial and final state fermions. We use the approximation that the
electron is massless, however as we have angular cuts to keep the
photons separated from the electrons we do not work in an area of
phase space where the electron mass terms are important.
We include both s channel $Z^0$ and photon
exchange.

For the physical constants we use,
$$
\eqalign{
\alpha_{\rm em}^{\rm int} &= 1/128 \cr
\alpha_{\rm em}^{\rm ext} &= 1/137 \cr
\sin \theta_W &= 0.23 \cr
M_{Z^0} &= 91.175 \GeV \cr
m_\mu &= 105.6584 \MeV \cr}
\tag
$$
where we use $\alpha_{\rm em}(M_Z)\equiv\alpha_{\rm em}^{\rm int}$ as
the coupling for the s channel $Z^0$ and photon, and
$\alpha_{\rm em}(0)\equiv \alpha_{\rm em}^{\rm ext}$ as the coupling
for all external photons \refto{BPQCD}.

As we have not removed the large logs associated with collinear
radiation in the Monte Carlo we are required to keep
away from the regions of phase space
where these logs become important, this means that we must keep all
photons (be they exponentiated or not) separate from all charged
particles. This we do by imposing the following experimental cuts,
$$
\eqalign{
\theta_{\mu\gamma} &> 5 ^\circ \cr
|\cos\theta_\gamma| &< 0.9 \cr} \tag colcut
$$
where $\theta_{\mu\gamma}$ is the angle between a muon and photon, and
$\theta_\gamma$ is the angle of a photon from the beam pipe.
In order for the final state particles to be observed we impose,
$$
\theta_{\mu\mu} > 20 ^\circ
$$
$$
35^\circ < \cos\theta_\mu < 145^\circ \tag
$$
where $\theta_{\mu\mu}$ is the angle between the two muons and
$\theta_\mu$ is the angle of a muon from the beam pipe.
We choose,
$$
\sqrt s = M_Z \tag
$$
Now for large $n$ the hard matrix element for process \(eemumupr) can
be very time consuming to calculate, so we calculate up to 4 photons
exactly -- then for larger numbers of photons we calculate the 4 most
energetic photons exactly, and then use the soft approximation to
calculate the remaining photons. In practice the contribution to the
cross-section from 4 or more hard photons is always very small, and
usually several of the photons are very soft; this means that we make
only a small error by using this approximation, while speeding up the
code considerably.
\topinsert
\centerline{
\epsfxsize=3 in \epsfbox[50 45 600 475]{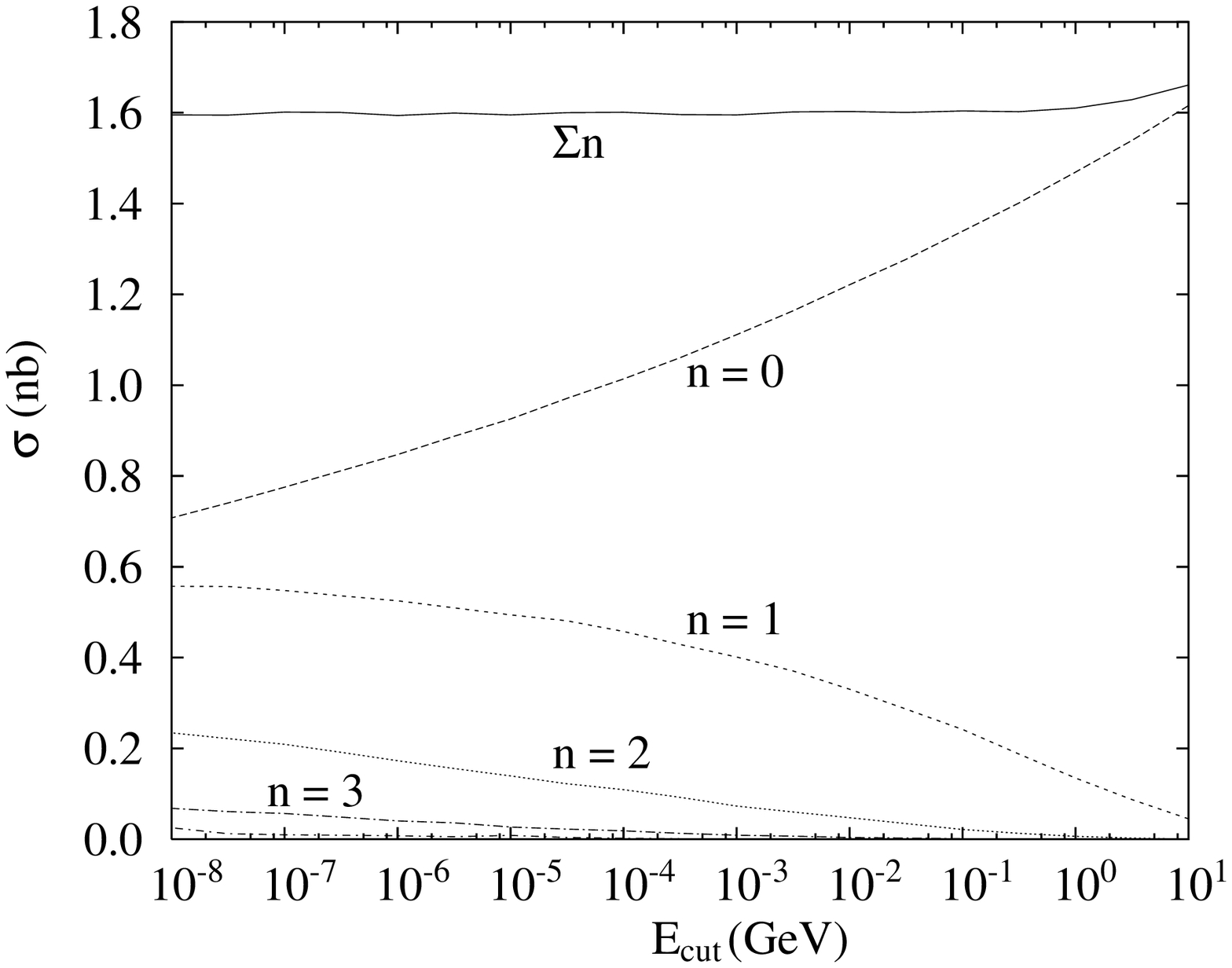}
\qquad
\epsfxsize=3 in \epsfbox[50 45 600 475]{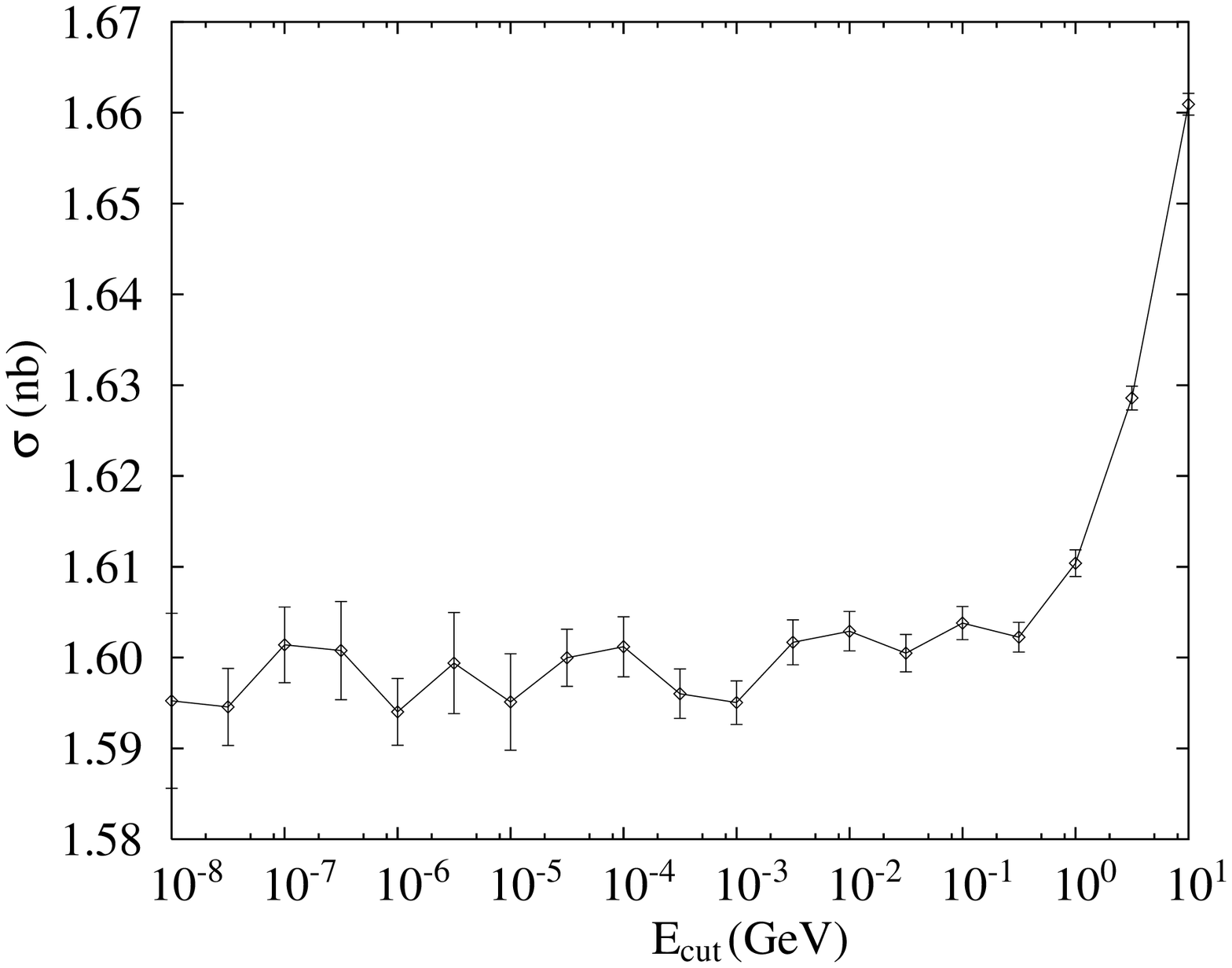}
}
\fig{ecut}{The contributions from varying numbers of hard photons to
the inclusive cross-section for the process
$e^+ e^- \to \mu^+ \mu^-$ as a function of $E_{\rm cut}$. We choose
$E_{\rm reg} = 10 \GeV$, $\sqrt s = M_Z$ and use the other cuts
described in the text.
On the right we expand the scale and just show
the inclusive cross-section. We also show the Monte Carlo errors
associated with each point.}
\endinsert
The first test of the program is the dependence upon $E_{\rm cut}$ and
$E_{\rm reg}$. In \rfig{ecut} we show the contributions from the $n$
hard photon cross-sections to the total inclusive cross-section as a
function of $E_{\rm cut}$. Clearly these each individually have a
large dependence upon $E_{\rm cut}$. However when we sum the
contributions from each number of photons to form the physical total
inclusive cross-section where photon radiation satisfies the cut
$\Omega_c$ we obtain the solid line which clearly has far
less dependence upon $E_{\rm cut}$. We also show the
total inclusive cross-section with an expanded scale. It is clear that
for small $E_{\rm cut}$ the cross-section is independent of $E_{\rm
cut}$ within the Monte Carlo error; however there is also a clear rise
in the total inclusive cross-section for values of
$E_{\rm cut} \gsim 1\GeV$. This can be understood as the soft
approximation that we have used to derive the exponential factor
\(expfac) breaks down for photon energies larger than
${\cal O}(1 \GeV)$.

As we take $E_{\rm cut}$ smaller it takes a longer time to calculate
the cross-section with the same accuracy as for smaller $E_{\rm cut}$
values we have a larger contribution from larger numbers of hard
photons; and the cross-sections for large numbers of photons are very
time consuming to calculate due to the rapid increase in the number of
Feynman diagrams. We have also calculated the $E_{\rm cut}$ dependence
of the more exclusive 1 and 2 photon cross-sections, and for small
$E_{\rm cut}$ values there is no visible dependence upon $E_{\rm cut}$.
This gives us confidence that our computer code has no mistakes in it.
In all the results that follows we will choose $E_{\rm cut} = 10^{-2} \GeV$
which is safely in the region where our results are independent of
$E_{\rm cut}$, also the value of $E_{\rm cut}$ is not so small that we
spend large amounts of time calculating the cross-section for large
numbers of hard photons.
\topinsert
\centerline{
\epsfxsize=4.5 in \epsfbox[50 45 600 475]{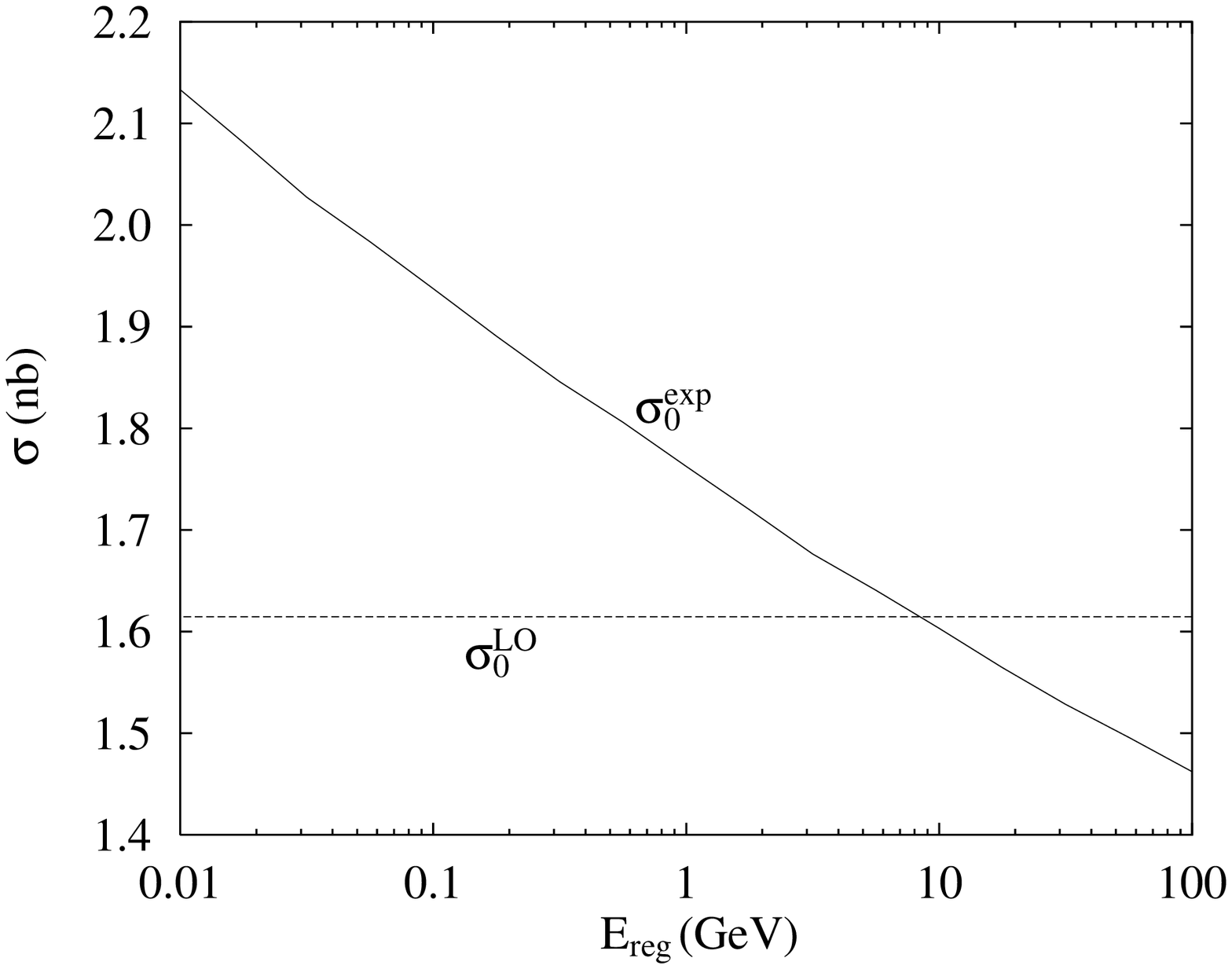}}
\fig{ereg}{The dependence of the inclusive $e^+e^- \to \mu^+\mu^-$
exponentiated cross-section on $E_{\rm reg}$. We also show the LO result
for the same cross-section.}
\endinsert
If \rfig{ereg} we show the dependence of the inclusive
cross-section upon $E_{\rm reg}$, and also the LO
$e^+e^- \to \mu^+ \mu^-$ cross-section. The dependence of the
cross-section on $E_{\rm reg}$ is clear to see.
The two cross-sections are not
directly comparable as the exponentiated result has dependance upon
$\Omega_c$; however it is clear that the cross-sections are comparable
as we expect, and the difference is of ${\cal O}(\alpha)$ for values
of $E_{\rm reg}\simeq 10\GeV$ as suggested by equation \(Ereg).
In all following results we will choose $E_{\rm reg} = 10\GeV$.
\topinsert
\centerline{
\epsfxsize=4.5 in \epsfbox[65 45 600 475]{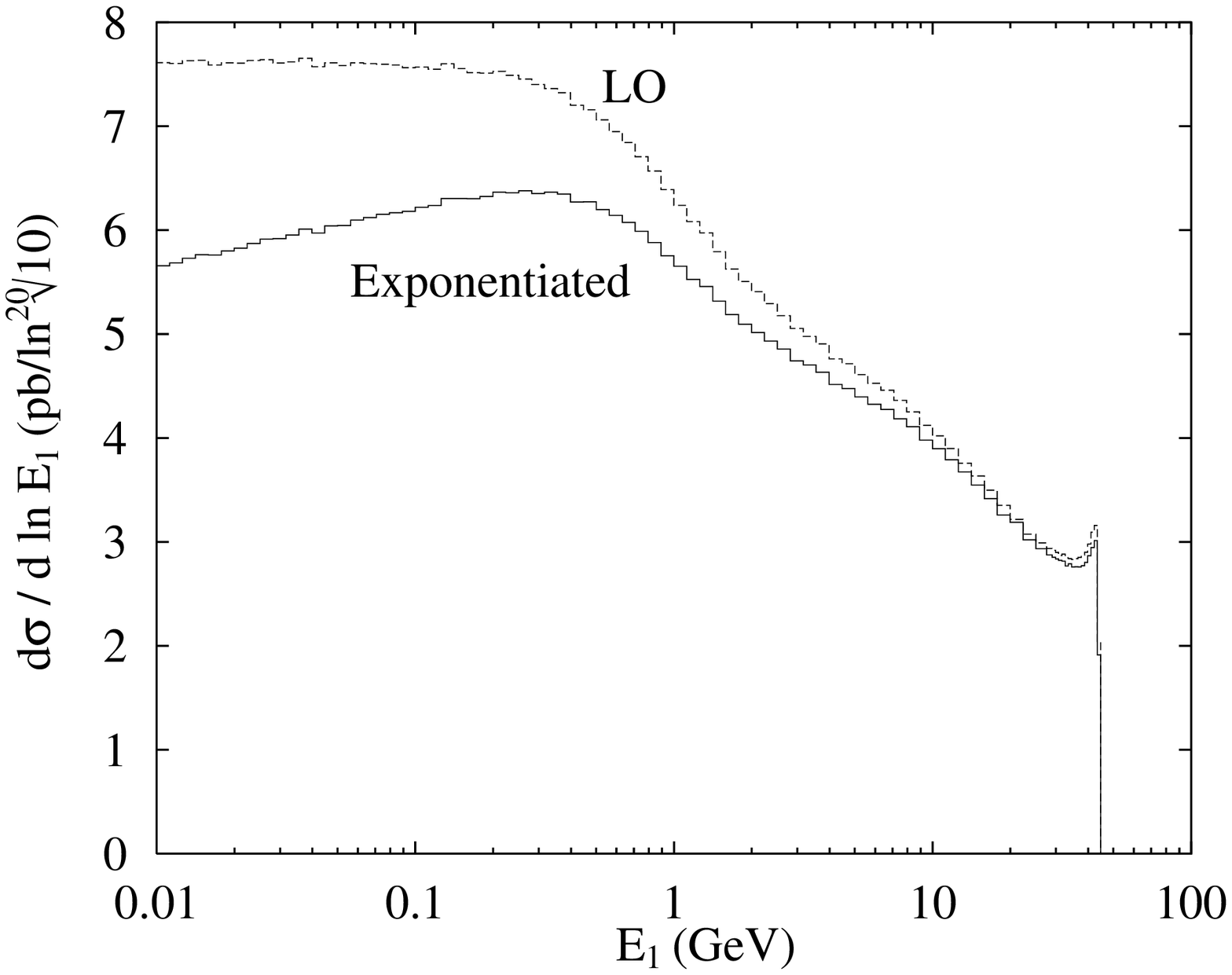}}
\fig{dsdlne1}{The 1 photon differential cross-section
$d\sigma / d \ln E_1$ plotted both at LO and exponentiated.}
\endinsert
Moving onto the more exclusive 1 photon differential cross-section, if
we consider events with 1 or more photons present then the energy of
the most energetic photon is a physical quantity. This means that we
can form the differential cross-section $d\sigma / d \ln E_1$, where
$E_1$ is the energy of the most energetic photon, in a meaningful way.
If we do this we find the results shown in \rfig{dsdlne1}; in this
figure we have also plotted the LO prediction for
$d\sigma / d \ln E_1$. As $E \to 0$ we see that the
$d\sigma^{\rm LO} / d \ln E_1 \to \hbox{\it constant}$; this is
manifest from the soft photon approximation. Equation \(sigma1) tells
us that,
$$
{d \sigma_1^{\rm LO} \over d \ln E_1 } =
  \sigma_0 {\alpha\over 4\pi^2} g(\Omega_c)
  \approx \hbox{\it constant} \tag
$$
That this breaks down for $E_1 \gsim 1 \GeV$ we can understand as
being due to initial state radiation (ISR), the large cross-section
comes from the process $e^+ e^- \to Z \to \mu^+ \mu^-$ with an extra
photon radiated; however if the photon is radiated from the initial
state with an energy greater that ${\cal O}(\Gamma_Z)$ this forces the
exchanged $Z^0$ boson far off mass shell, and this suppresses the
cross-section.

We can see that for large $E_1$ the LO 1 photon cross-section gives an
answer similar to the exponentiated 1 photon cross-section; however
for smaller $E_1$ the LO result over estimates the cross-section. The
effect of resumming the logarithms associated with the soft
singularity is to change the shape of the 1 photon cross-section. Notice
that changing the value of $E_{\rm reg}$ largely just changes the
overall normalisation by a factor of,
$$
(E_{\rm reg})^{- \alpha g(\Omega_c)/4 \pi^2} \tag
$$
and does not change the overall shape of the 1 photon cross-section.

For $E_1 \simeq \sqrt s / 2$ the cross-section has a small maximum,
this is mainly due to $e^+ e^- \to \mu^+ \mu^-$ production followed by
a muon fragmenting into a photon. It is clear to see why this
produces a maxima for large photon energies, consider,
$$
\mu^*(1) \to \mu(2)\ \gamma(3) \tag mufrag
$$
this fragmentation is proportional to the propagator of muon 1; \ie it
is inversely proportional to $p_1^2 = (p_2+p_3)^2$, the final muon
photon invariant mass. If we look at a fragmentation at a particular
$p_1^2 = (p_2+p_3)^2$ value, if the photon and muon 2 have similar
energies then they are produced largely collinear and the cut \(colcut)
vetos this event. However if either the final muon, 2, or the photon
is very soft then the angle $\theta_{\mu\gamma}$ is far larger and we
pass the collinear cut \(colcut). The configuration where the photon
carries most of the energy of muon 1 gives rise to the small peak in
\rfig{dsdlne1} at large $E_1$.
\topinsert
\centerline{
\epsfxsize=4.5 in \epsfbox[65 45 600 475]{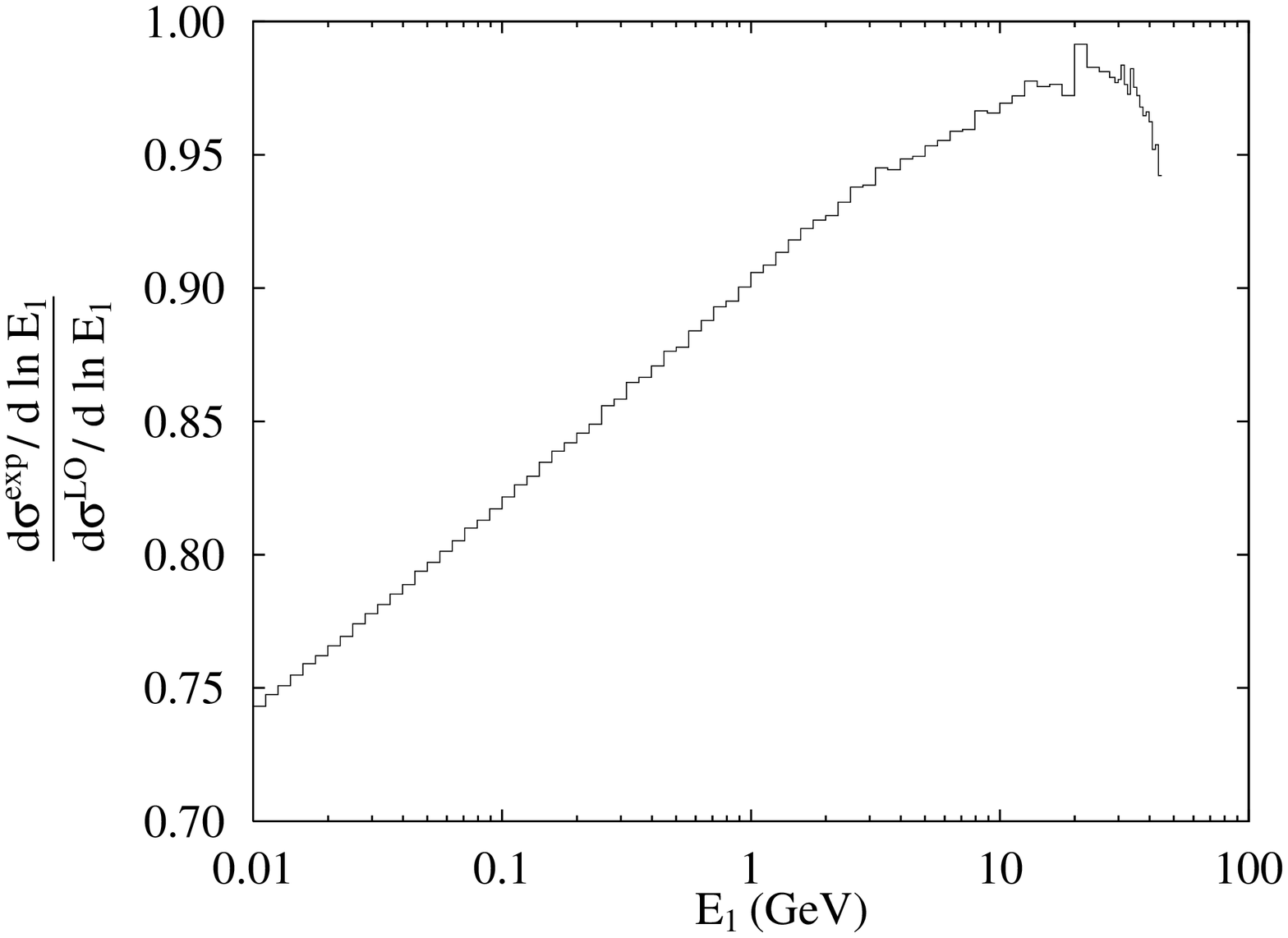}}
\fig{e1ratio}{The ratio of the exponentiated divided by the LO 1
photon differential cross-section  as a function of the photon energy,
$E_1$.}
\endinsert
In \rfig{e1ratio} we plot the ratio,
$$
{ d\sigma^{\rm exp}/d \ln E_1 \over d\sigma^{\rm LO}/d \ln E_1} \tag
$$
We can see that there is no particular energy where exponentiation
becomes important, rather thats as we consider lower and lower energy
photons exponentiation slowly becomes a more important factor. For our
choice of $E_{\rm reg} = 10\GeV$ the exponentiated result is comparable
to the LO results for large $E_1$ (the dip close to $\sqrt s /2$ is
due to the the process \(mufrag) mentioned before), however for photon
energies of $1\GeV$ the exponentiated result is down to 90\% of the LO
result.

It is of interest to ask how the soft photon approximation compares to
the exact matrix element result.
However before we compare the two approaches there is a
subtlety with the soft photon approximation that should be
appreciated. The soft photon approximation tells us that the matrix
element with a photon present is the matrix element without that
photon multiplied by an eikonal factor. However what does the matrix
element without the photon mean? Consider the process
$e^+ e^- \to \mu^+ \mu^- \gamma$, if we remove the photon then we no
longer have 4 momenta conserved; in particular if we ask what is the
$Q^2$ carried by the $s$ channel $Z^0$ or $\gamma$ we can form this in
two ways as $(p_{e^+}+p_{e^-})^2$ or as $(p_{\mu^+}+p_{\mu^-})^2$  and
these have different values. As such the soft photon
approximation has ambiguities associated with what the matrix element
without photons is. Clearly as the photon energy goes to zero this
ambiguity disappears.

In the current case of $e^+ e^- \to \mu^+ \mu^-$ with $\sqrt s = M_Z$
this ambiguity can be of great importance for photon energies greater
than ${\cal O}(\Gamma_Z)$. If we have initial state radiation (ISR)
with energy greater than ${\cal O}(\Gamma_Z)$ then this forces the
exchanged $Z^0$ boson off mass shell, and this strongly suppresses the
cross-section; this means that the relevant $Q^2$ for the exchanged
$Z^0$ is $(p_{\mu^+}+p_{\mu^-})^2$ which means that for photon
energies greater than ${\cal O}(\Gamma_Z)$ the $Z^0$ propagator is
strongly suppressed. Similarly if we have final state radiation (FSR)
with energy greater than ${\cal O}(\Gamma_Z)$ then the relevant $Q^2$
is $(p_{e^+}+p_{e^-})^2$. With this in mind, and given that photon
radiation is radiated off a pair of charged legs (at least for massless
fermions) a good prescription is to boost the two legs that the photon
can be radiated off in such a way that 4 momentum is conserved and
that their momentum only changes by a small amount. In the present
case this is messy as the soft eikonal factor \(eikonal) has 6
different combinations of legs that it can be radiated off, and so
there are 6 different boosts that need to be carried out. This is
beyond the scope of the current work; instead we consider the two
simpler cases of pure ISR and pure FSR (\ie the cases where we take
the final and initial legs respectively to be chargeless). For pure ISR
(FSR) it is clear which legs the photon is radiated off, and hence
which legs should be boosted; for ISR we boost the electron legs, and
for FSR we boost the muon legs. If we are to boost momenta $p_1$ and
$p_2$ we use the boost defined by,
$$
\eqalign{
p_1' &= a_{11}\, p_1 + a_{12}\, p_2 + p_T \cr
p_2' &= a_{21}\, p_1 + a_{22}\, p_2 + p_T \cr
}\tag
$$
with $p_1'^2=p_1^2$, $p_2'^2=p_2^2$, $\ddot{p_1}{p_T}=0$ and
$\ddot{p_2}{p_T}=0$; where we require $p_1'$ and $p_2'$ to satisfy 4
momentum in the system without photons.
%
\topinsert
\centerline{
\epsfxsize=3 in \epsfbox[30 50 595 470]{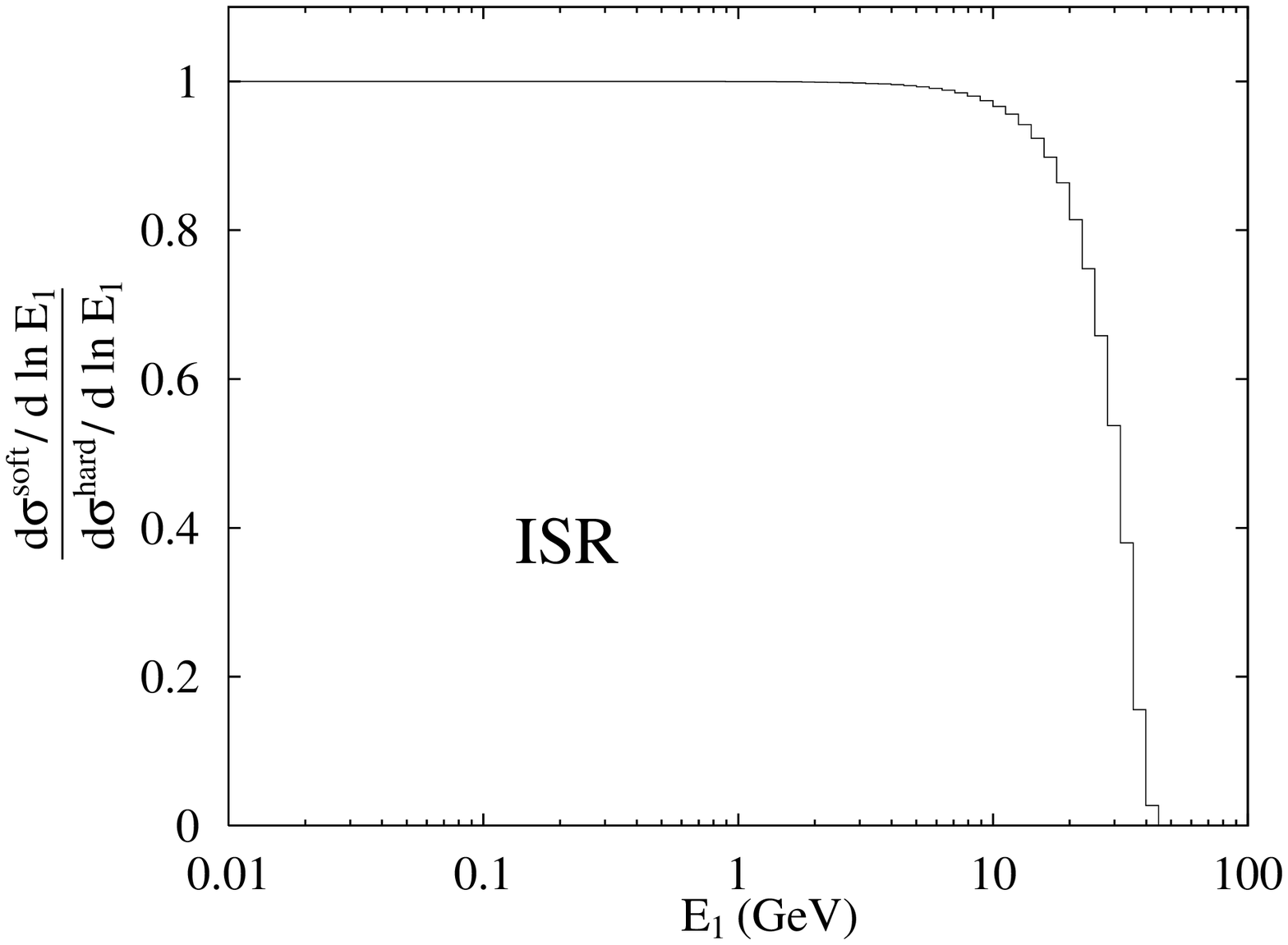}
\qquad
\epsfxsize=3 in \epsfbox[30 50 595 470]{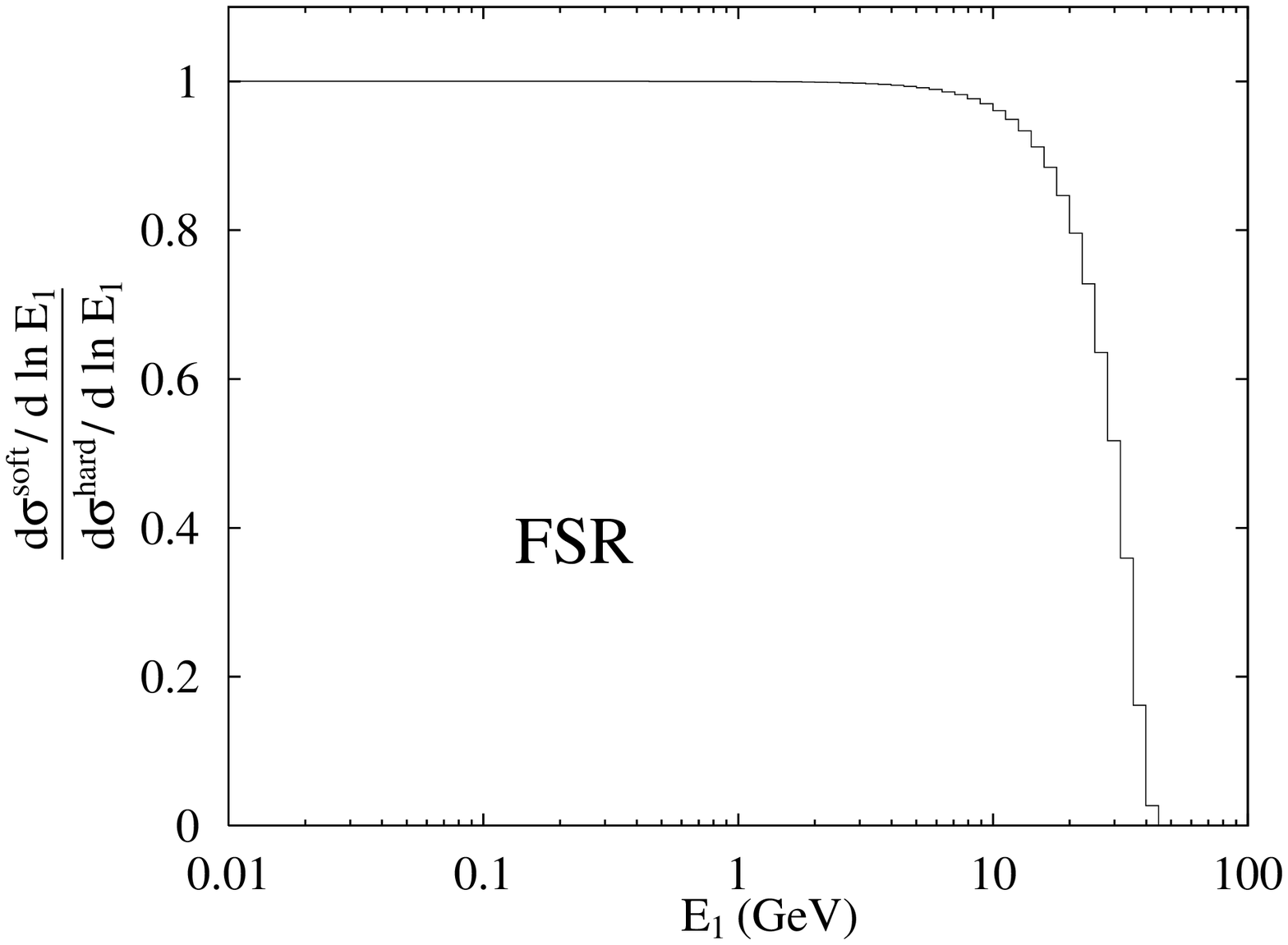}
}
\fig{ratiohs}{The soft photon approximation compared to the hard
matrix element, both exponentiated, for the cases of pure ISR and pure
FSR.}
\endinsert
We plot the ratio of the soft approximation to the hard matrix element
in \rfig{ratiohs}. It is clear that the soft approximation is good for
photon energies up to ${\cal O}(10\GeV )$ but underestimates the hard
result for larger energies. For the case of ISR this makes little
difference as radiation with energy larger than ${\cal O}(\Gamma_Z)$
is strongly suppressed; and so we only make mistakes where the
contribution to the cross-section is small. The rate for hard FSR can be
quite appreciable and for this hard radiation the soft approximation
underestimates the cross-section by a large factor. It seems hard to
improve the implementation of the soft photon approximation in such a
way that it correctly describes these hard photons.

As a final numerical result we wish to consider the photon photon
invariant mass distribution in the process,
$$
e^+ e^- \to \mu^+ \mu^- \gamma \gamma \tag
$$
this process has been the subject of much interest since the L3
collaboration at LEP announced 4 events with $M_{\gamma\gamma} = 60
\GeV$ \refto{L3hardphot} in apparent conflict with the Standard
Model. Although the conflict with the Standard Model has since gone
away with improved statistics this process serves as a good
illustration of a process that contains both soft and hard photons and
so both exponentiation of soft photons and exact matrix elements for
hard photons are important.
\topinsert
\centerline{
\epsfxsize=4.5 in \epsfbox[50 50 595 480]{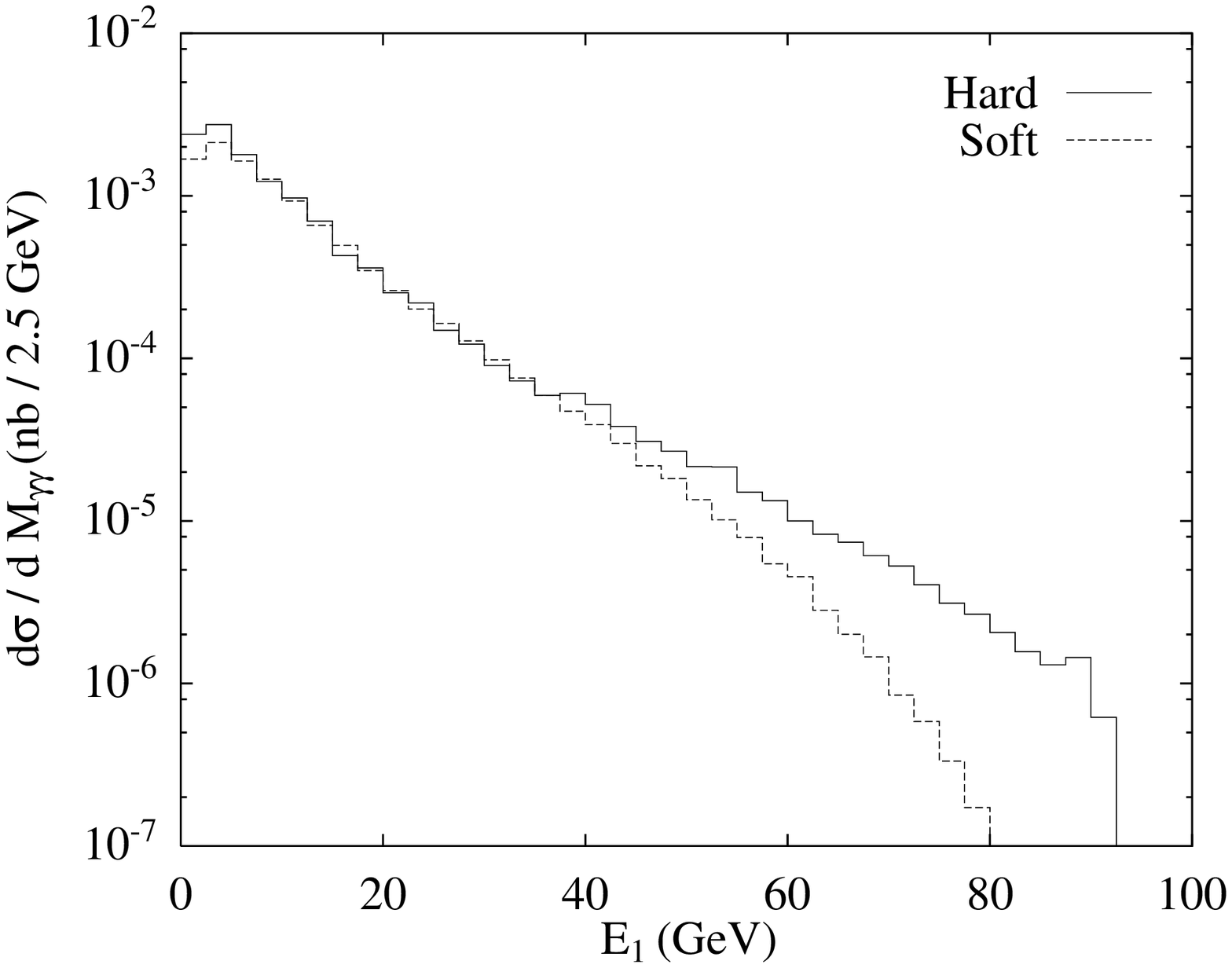}}
\fig{dsdmff}{The exponentiated differential cross-section
$d\,\sigma / d\,M_{\gamma\gamma}$ both with hard matrix elements
and in the case where the hardest photon is treated exactly and the
remaining photons are described by the soft photon approximation.}
\endinsert
We use cuts similar to the L3 experimental cuts;
$$
\eqalign{
|cos \theta_\gamma | < 0.9 \qquad & \qquad
            36^\circ < \theta_\mu < 144^\circ \cr
\theta_{\gamma\mu} > 5^\circ \qquad & \qquad
            \theta_{\mu\mu} > 20^\circ \cr
E_\gamma &> 1 \GeV \cr
}\tag
$$
We show the exponentiated hard differential cross-section we obtain in
\rfig{dsdmff}. Now we can consider various different approximations to
this process. If we just use the LO hard matrix element result we
would expect to describe large $M_{\gamma\gamma}$ well, but have
larger difficulty with small $M_{\gamma\gamma}$; in practice with the
cuts used there is no difference within the Monte Carlo errors from
the exponentiated hard result for all values of $M_{\gamma\gamma}$.

We also compare the soft photon approximation, for the comparison we
have boosted the final state muons in order to conserve 4 momenta
within the soft approximation. This keeps the exchanged $Z^0$ on mass
shell even when we have hard ISR, as such we would expect this to
overestimate the cross-section. We show the exponentiated
cross-section where we calculate the hardest photon using the exact
matrix element and calculate additional photons using the soft
approximation. It is clear that although this works well for small
$M_{\gamma\gamma}$ this {\it underestimates} the cross-section by a
large amount for large $M_{\gamma\gamma}$. Surprisingly the
differential cross-section is almost unchanged when we calculate all
photons within the soft approximation from the result when we treat
one photon exactly.

For this process it is clear that describing the hard photons
correctly is more important than exponentiating the soft photons.

\head{Conclusions}

In this paper we have presented a method for exponentiating the
singularities associated with soft photon emission while still
retaining exact matrix elements for hard photon emission. We have
applied this method to the process,
$$
e^+ e^- \to \mu^+ \mu^- + n \gamma \tag ceemumu
$$
where the photons can be radiated from any of the charged legs, \ie we
have included ISR, FSR and all the interference terms. We have also
retained the mass dependence of the muon.

Exponentiation is of importance for low energy photons where it
suppresses the cross-section over the LO prediction, whereas exact matrix
elements are important for hard photons when the soft approximation
that is the basis of exponentiation breaks down. Experimentally an
accurate description of both soft and hard photons is often
simultaneously required as in a process where the hard photons are of
primary interest the low energy photons are used to normalise the
theoretical predictions for that process.

We have give a method by which one could also remove the collinear
singularities, and resum the associated logarithms, in section 3.
As we are not currently interested in these singularities we have made
no attempt to include these effects in our Monte Carlo; but instead
impose experimental angular cuts on all photons.

\headnn{Acknowledgements}

I would like to thank Nigel Glover and Walter Giele for most useful
conversations. I would also like to thank the Theory Division of
FermiLab for its kind hospitality where part of this research was
carried out.

This research was supported in part by SERC in the form of a research
studentship, in part by the University of Wisconsin Research
Committee with funds granted by the Wisconsin Alumni Research Foundation, in
part by the U.S.~Department of Energy under Contract
Nos.~DE-AC02-76ER00881 and DE-FG02-95ER40896,
and in part by the Texas National Research Laboratory Commission under
Grant No.~RGFY93-221.

\vfill\eject
\headnn{Appendix}
\taghead{A.}

In this appendix we give the explicit form of the function
$g(\Omega_c)$. $g$ is defined as,
$$
g(\Omega_c)=\int_{\Omega_c} f(\Omega) d \Omega \tag
$$
with,
$$
f(\Omega) = E^2
   \left( {p_1 \over \ddot{p_1}{k}} - {p_2 \over \ddot{p_2}{k}} +
        {p_3 \over \ddot{p_3}{k}} - {p_4 \over \ddot{p_4}{k}} \right)^2
\tag
$$
Now $f(\Omega)$ contains two types of term,
$E^2 \ddot{p_1}{p_2}/\ddot{p_1}k \ddot{p_2}k$ and
$E^2 m^2 /(\ddot{p_1}k)^2$. Concentrating on the first of
these we have to integrate,
$$
\eqalign{
I_1 &=
\int 2 E^2 {\ddot{p_1}{p_2}\over\ddot{p_1}k \ddot{p_2}k} d\Omega \cr
&= \int {2 (E_1 E_2 - P_1 P_2 \cos\rho)\ d \cos\theta\ d \phi \over
(E_1-P_1 \cos\theta)
(E_2-P_2\cos\rho \cos\theta - P_2 \sin\rho \sin\theta \cos\phi)} \cr}
\tag
$$
where we have written,
$$
\eqalign{
p_1 &= (E_1,P_1,0,0) \cr
p_2 &= (E_2,P_2 \cos \rho, P_2 \sin\rho, 0) \cr
k   &= (E , E\cos\theta , E\sin\theta\cos\phi , E \sin\theta\sin\phi) \cr}
\tag
$$
Integrating over $\phi$ gives,
$$
I_1 = \int { 2\pi \ d\cos\theta \over
      ( E_1 - P_1\cos\theta) ((P_2 \cos\theta - E_2 \cos\rho)^2 +
m_2^2\sin^2\rho)^{1/2}}
\tag
$$
and performing the indefinite integral over $\cos\theta$,
$$
\eqalign{
I_1(p_1,p_2,\cos\theta )&=
	{2\pi \over (z^2 + m_2^2\sin\rho^2\, P_1^2)^{1/2}} \times\cr
& \qquad \ln \left({ (y^2+m_2^2\sin^2\rho)^{1/2}
                (z^2 +m_2^2\sin^2\rho\, P_1^2)^{1/2}
  +zy +m_2^2\sin^2\rho\, P_1 \over E_1-P_1\cos\theta}\right) \cr}
\tag gint
$$
where $y$ and $z$ are defined as,
$$
\eqalign{
y &= -P_2 \cos\theta -E_2 \cos\rho \cr
z &= P_2 E_1 - E_2 p_1 \cos\rho \cr}
\tag
$$
Now we want integrate $f(\Omega)$ over the region defined by
$\Omega_c$, that is the whole of space with small angular regions
removed about the charged particles. Two of those charged particles are
$p_1$ and $p_2$, now $\theta$ is the angle between $p_1$ and the
photon, and so we can form the integration over the region with
$\theta_{1\gamma} > \theta_1$ and $\theta_{2\gamma} > \theta_2$ as,
$$
I = (I_1(p_1,p_2,\cos\theta_1) - I_1(p_1,p_2,-1))
   - (I_1(p_2,p_1,1) - I_1(p_2,p_1,\cos\theta_2))
\tag ggint
$$
Now as $m_2\to 0$, $I_1(p_1,p_2,-1)$ and $I_1(p_2,p_1,1)$ diverge, and
these two divergences cancel on each other. In order to avoid
numerical difficulties it is wise to extract the singular terms and
cancel them by hand. This we can do by rewriting the logarithm in
equation \(gint) as,
$$
\eqalign{
&\ln \left( (E_1+P_1){(y^2+m_2^2\sin^2\rho)^{1/2}
                (z^2 +m_2^2\sin^2\rho\, P_1^2)^{1/2}
  +zy +m_2^2\sin^2\rho\, P_1 \over
 m_1^2+(1-\cos\theta)P_1(E_1+P_1) }\right) \qquad\cr
& \quad = \ln \left( {m_2^2\sin^2\rho \over E_1-P_1\cos\theta}
\left( {z^2+y^2 P_1^2 + m_2^2\sin^2\rho\,
P_1^2 \over (y^2+m_2^2\sin^2\rho)^{1/2}
    (z^2 +m_2^2\sin^2\rho\, P_1^2)^{1/2} - zy } + P_1 \right)\right)
 \cr}
\tag
$$
In these forms we can explicitly see the $\ln(m_2^2)$ term and as
$\cos\theta\to 1$ the $-\ln(m_1^2)$  term and cancel them by hand.

So far we have assumed that the angular cuts about charged particles
do not overlap, in practice if the angular cuts overlap the charged
particles are close in phase space and so $\ddot{p_1}{p_2}$ is small
and the eikonal factor as a whole is small. Numerically the problem
that we encounter is that $I$ goes negative, in the case that this
happens we set $I=0$ and as the eikonal factor is small this only
makes a small error.

In addition to particles 1 and 2 we also have two additional particles 3
and 4 that the photon is kept separated from. However these particles
are typically far away from the singularities of $f$ and is slowly
varying. This means we can
approximate the effect of the cuts $\theta_{3\gamma}> \theta_3$ and
$\theta_{4\gamma}> \theta_4$ by evaluating
$E^2 \ddot{p_1}{p_2}/\ddot{p_1}k \ddot{p_2}k$ in the
direction of particles 3 and 4 and then multiplying this by the volume
of space that the cut excludes,
$$
2\pi ( 1 - \cos\theta_3)
\left( {E^2 \ddot{p_1}{p_2}\over\ddot{p_1}k \ddot{p_2}k} \right)
\Bigg|_{k \| p_3}
\qquad \hbox{and}
\qquad 2\pi ( 1 - \cos\theta_4)
\left( {E^2 \ddot{p_1}{p_2}\over\ddot{p_1}k \ddot{p_2}k} \right)
\Bigg|_{k \| p_4}
\tag
$$
we then subtract these two contributions from \(ggint).

It remains to calculate the integral of $E^2 m^2 /(\ddot{p_1}k)^2$.
This is given by
$$
\eqalign{
I_2(\cos\theta) &\equiv \int {E^2 m^2 \over (\ddot{p_1}k)^2}\ d\Omega
= \int d\cos\theta \int_0^{2\pi} d \phi {m^2
               \over (E_1-P_1\cos\theta)^2} \cr
&= {2\pi\, m^2 \over P_1 ( E_1 - P_1 \cos\theta )} \cr
}
\tag i2
$$
As before we impose the angular cut about particle 1 by integrating \(i2)
between $-1 < \cos\theta < \cos\theta_1$, and impose the angular cuts
about particles 2,3 and 4 by evaluating $E^2 m^2 /(\ddot{p_1}k)^2$ in
the directions of particles 2,3 and 4 and multiplying by the volume of
the angular cut.

\vfill\eject
\references

\refis{YFS} D.R. Yennie, S. Frautschi, and H. Suura, {\it Ann. Phys}
{\bf 13} (1961) 379,\hfil\break
K.T. Mahanthappa, {\it Phys. Rev.} {\bf 126} (1961) 329.

\refis{Hard}
W.J. Stirling {\it Phys.Lett.B} {\bf 271} (1991) 261.

\refis{Soft}
F.A. Berends, R. Kleiss, and S. Jadach, {\it Comp. Phys. Commun.}
{\bf 29} (1983) 185.

\refis{Collinear}
J.E. Campagne and R. Zitoun, An expression of the electron structure
function in QED, LPNHEP-88-06 (1988). \hfil\break
J.E. Campagne and R. Zitoun, {\it Z.Phys} {\bf C43} (1989)
469.\hfil\break
G. Bonvicini and L.Trentadue, {\it Nucl. Phys.} {\bf B323} (1989)
253.\hfil\break
J. Fujimoto,  Y. Shimizu, and T. Munehisa, KEK-PREPRINT-92-193 (1992).

\refis{JW}
S. Jadach and B.F.L Ward, {\it Comp.Phys.Comm.} {\bf 56} (1990) 351.

\refis{YFS3}
S. Jadach and B.F.L Ward, {\it Phys. Lett.B} {\bf 274} (1992) 470.

\refis{BHLUMI4}
``BHLUMI4.00: YFS Monte Carlo Approach to high precision low angle Bhabha
scattering at LEP / SLC energies.''
S. Jadach, E. Richter-Was, B.F.L. Ward, Z. Was,UTHEP-94-0602.

\refis{WWapprox} C. von Weizs\"acker, {\it Z. Phys.} {\bf 88} (1934)
612,\hfill\break
E. Williams, {\it Phys. Rev.} {\bf 45} (1934) 729.

\refis{L3hardphot} The L3 Collaboration, {\it Phys.Lett.B} {\bf 295} (1992)
337.

\refis{BPQCD} See sec 1.2.5 of `Basics of Perturbative QCD', Yu.L.
Dokshitzer, V.A. Khoze, A.H. Mueller, and S.I. Troyan, Editions
Fronti\`eres (1991).

\refis{YRIII} R. Kleiss \etal , `$Z$ Physics at LEP1', CERN Yellow
Report 89-08 (1989) vol.3, page 1.

\refis{mmff} D.J.Summers, {\it Phys.Lett.} {\bf B302} (1993) 326.

\refis{KLN} T. Kinoshita and A. Sirlin, {\it Phys.Rev.} {\bf 113}
(1959) 1652,\hfill\break
T. Kinoshita, {\it J.Math.Phys.} {\bf 3} (1962) 650,\hfill\break
T.D. Lee and M. Nauenberg, {\it Phys.Rev.} {\bf 133} (1964) 1549.

\refis{BN} F. Bloch and A. Nordsieck, {\it Phys.Rev.} {\bf D52} (1937) 54.

\endreferences

\endit